Title Page

# Value-at-Risk forecasting model based on normal inverse Gaussian distribution driven by dynamic conditional score

Shijia Song
202131250022@mail.bnu.edu.cn
School of Systems Science, Beijing Normal University, Beijing 100875, China

Handong Li
lhd@bnu.edu.cn
School of Systems Science, Beijing Normal University, Beijing 100875, China

Correspondence should be addressed to Handong Li
E-mail: lhd@bnu.edu.cn
Tel. 8610-58807064
Tax. 8610-58807876

# Value-at-Risk forecasting model based on normal inverse Gaussian distribution driven by dynamic conditional score

## Abstract

Under the framework of dynamic conditional score, we propose a parametric forecasting model for Value-at-Risk based on the normal inverse Gaussian distribution (Hereinafter NIG-DCS-VaR), which creatively incorporates intraday information into daily VaR forecast. NIG specifies an appropriate distribution to return and the semi-additivity of the NIG parameters makes it feasible to improve the estimation of daily return in light of intraday return, and thus the VaR can be explicitly obtained by calculating the quantile of the re-estimated distribution of daily return. We conducted an empirical analysis using two main indexes of the Chinese stock market, and a variety of backtesting approaches as well as the model confidence set approach prove that the VaR forecasts of NIG-DCS model generally gain an advantage over those of realized GARCH (RGARCH) models. Especially when the risk level is relatively high, NIG-DCS-VaR beats RGARCH-VaR in terms of coverage ability and independence.

**Keywords**

# 1  Introduction

The wide application of electronic trading systems in the financial market and the massive increase in the number of assets traded by financial institutions have made risk measurement and risk control the main focuses of financial regulatory agencies. According to the Basel Accord proposed by the Bank for International Settlements (BIS) in 1996, the risk capital of the bank must be sufficient to make up for 99% of the possible losses during the 10-day holding period. This value is so called value-at-risk (VaR). In actual supervision, the holding period and confidence level may vary from situation to situation. Since then, VaR has become the most popular risk management tool in the financial market. It represents the estimated maximum loss of an asset over a specified holding period and a specified level of probability (Alexander, 2008). Although Artzner et al. (1997) clearly stated that VaR has various theoretical flaws and may fail to accurately measure market risks, VaR is still one of the most important risk management tools. Therefore, looking for improved VaR estimating methods and improving the accuracy of risk forecast have become the core issue in relative fields.

Returns of stock prices are usually sampled by short time intervals (daily or weekly) and show non-normality. It is well known that the empirical distribution of returns tends to have sharp peak and fat tails, indicating a higher frequency of extreme returns than under the normal assumption. Additionally, the distribution is often left-skewed, which suggests that large losses occur more frequently than large profits. Therefore, from the perspective of risk management, the left tail of returns deserves more attention.

The NIG distribution was introduced by Barndorff-Nielsen (1997), which can model symmetric or asymmetric distributions that may have heavy tails in both directions. Moreover, the NIG distribution also has some special theoretical properties, e.g., it is closed under affine transformation and convolution. The analytical tractability of NIG makes it possible to model financial returns in wide applications. Andersson (2001), Jensen and Lunde (2001), Forsberg and Bollerslev (2002),Venter and de Jongh (2002) specified NIG as the conditional distribution of the GARCH models. Eberlein and Keller (1995), Prause (1997), Rydberg (1997), Bølviken and Benth (2000) and Lillestøl (2000) performed NIG as unconditional return distribution. The tail behavior of NIG is usually regarded semi-heavy, i.e., it has a heavier tail than the Gaussian distribution. But when fitting for returns with much more extreme values, NIG is not as good as non-Gaussian stable distributions such as Pareto distribution.

Barndorff-Nielsen (1997) further pointed out that the NIG distribution is closed under convolution in the following sense: if multiple independent variables are NIG-distributed with the same tail and skewness parameters, and possibly different location and scale parameters, then the sum of the independent variables will also be NIG-distributed with the same tail and skewness parameters. But its location and scale parameters are obtained by summing up the corresponding parameters of each independent variable separately. Since it is a new trend to incorporate high frequency (HF) information of intraday return to construct financial models, this property of NIG provides an innovative and tractable method for fitting distribution of daily return by directly using intraday returns.

Many researchers have made various efforts and considerations to improve the fitting accuracy of financial returns. One of the practices is to take the information of intraday returns into account and try to integrate their effect in the model to make a more accurate measurement of daily returns. Since Andersen and Bollerslev (1997), Andersen et al. (2001a), Andersen et al. (2001b), Barndorff-Nielsen and Shephard (2002) pioneered the use of realized volatility measures as an effective and consistent estimator of implied volatility, the availability of high-frequency intraday data has aroused extensive discussion and inspired a lot of research on the modeling of realized volatility and prediction of daily VaR. In the literature of the realized volatility model, many scholars have confirmed that it is significantly better than the ARCH-type volatility model (Andersen et al., 2003; Koopman et al., 2005; Martens et al., 2009). Andersen et al. (2001), Engle and Giampiero (2006) also highlighted the importance of intraday information in the studies of realized volatility. Bee et al. (2019) proposed a dynamic structure for exceedance modeling based on HF data by adding realized volatility to the POT model. Although it is not common to directly use HF data to improve the risk management model, currently more and more studies try to extend the models constructed originally based on low frequency (LF) data in this way. Shephard and Sheppard (2010), Noureldin et al. (2012), Hansen and Huang (2016) considered using HF data to fit the distribution of conditional return, further demonstrating the significance of HF data. Cai et al. (2019) implemented a functional autoregressive model for VaR forecasts by estimating the possibility density function of intraday returns. They proved that the model incorporated HF information could enhance the coverage of the daily VaR forecasts. Song et al.(2021) also constructed a VaR forecasting model combined with intraday returns and driven by dynamic conditional score. This parametric model assumed that the distribution of return follows a censored generalized Pareto distribution.

In addition to enriching the content of the parametric model for returns, another effort to enhance the precision of estimation is to find a robust and available tool to estimate the core parameters appropriately. In financial risk management, it is crucial to capture the motion law of the parameters that control the shape of returns. However, there is no clear standard for setting a certain parameter as static or dynamic and many empirical studies confirmed that it should depend on the specific situation. Some financial models (Engle, 1982; Bollerslev, 1986; Nelson, 1991; Harvey and Shephard, 1996; Barndorff-Nielsen and Shephard, 2002; Bee et al. 2019) assumed the scale parameter of the distribution is dynamic and the shape parameter is constant. But others (Massacci, 2017; Astrid and Szabolcs, 2019; Harvey and Ito, 2020) set all parameters included in the distribution to be dynamic. Those models with dynamic parameters can be divided into parameter-driven models and observation-driven models (Cox ,1981). The latter includes autoregressive conditional heteroscedasticity (ARCH) model (Engle,1982), generalized autoregressive conditional heteroscedasticity (GARCH) model (Bollerslev ,1986), and dynamic-conditional-score (DCS) model (Creal et al., 2012; Harvey, 2013). What they have in common is that the variation of parameters is driven by the function of the observations. Recently, DCS has become popular in solving dynamic parametric problems since its mechanism to update the parameters over time is the scaled score of the likelihood function (Creal et al., 2012). Based on this score-driven framework, Zhang and Bernd (2016), and Massacci (2017) extended two dynamic models to estimate the probability of extreme financial returns and the size of the exceedance, respectively. Astrid et al. (2019) proposed a DCS model based on the NIG distribution, which can simultaneously update the volatility through scale parameter and shape parameter. Encouraged by the fact that the DCS model is adaptable to time-varying data and has a higher efficiency of parameter estimation, we consider constructing models under its framework to estimate the dynamic parameters.

The main contributions of our study include: first, the method of using the conditional distribution function to estimate VaR considers the time-varying property of parameters, which means that not only the time-varying mean and variance could be determined, but also the time-varying property of each moment. Second, we extend the method of using intraday returns to infer the distribution of daily returns. According to the strong form of efficiency market hypothesis, moment-specific returns are independent from each other.

Therefore, we could assume that the conditional distribution of intraday returns follows the independent NIG distribution. This assumption makes it easy to simulate daily returns by directly adding up the intraday returns or to determine the parameters in distribution of daily returns in terms of parameters in individual distribution of intraday returns. However, high-frequency return series in reality often have correlation, which may be related to that they all follow the same intraday cycle structure. Therefore, we will consider adding a deterministic periodic factor to the DCS model to reduce the correlation of sequences as much as possible, so as to make this simulating method more reasonable. As the literatures indicate, using intraday data to estimate the distribution of daily returns makes sense due to the abundant information provided by high-frequency trading. Compared with modelling based on historical daily return, models incorporated intraday return can reflect more real risk fluctuation within a day, thereby giving more accurate VaR estimates and forecasts. Third, under the framework of DCS, we capture the dynamic evolution process of time-varying parameters in a more precise way and obtain their one-step forecasts so that the daily VaR forecast can be produced by calculating the quantile of the predicted distribution of daily return. Finally, we use this model to conduct an empirical analysis of the Chinese stock market, and compare the performance of the NIG-DCS-VaR model and the realized-GARCH-VaR model through several backtestings as well as the model confidence set (MCS) approach. The results confirm that our model gain an advantage in estimating the risk of tail returns, which can be seen as an effective contribution to risk management.

The remainder of the paper is organized as follows. Section 2 outlines our NIG-DCS-VaR model. It first gives the overview of NIG distribution, describes the mechanism of DCS model based on NIG distribution and its maximum-likelihood (ML) estimator. Then it introduces a parametric and a non-parametric methods of degerming the distribution of daily return based on intraday return. Finally it illustrates three classic test methods and MCS approach for measuring the effect of out-of-sample VaR forecasts. Section 3 details the stock data used in empirical analysis and the corresponding data processing. Section 4 where the test results indicates that the out-of-sample VaR forecasts obtained by the model containing HF data are less likely to underestimate risk than that from model constructed by daily returns, and NIG-DCS-VaR beats RGARCH-VaR in terms of coverage ability and independence at some risk level. Section 5 concludes our work. Some supplementary materials are relegated to the Appendix.

## 2 Methodology

## 2.1 NIG Distribution

The NIG distribution is essentially a special case of the GH family. The univariate GH distribution can be parameterized in a variety of ways. The more common expression of the probability density function of GH was introduced by Prause (1999), and can be written as:

$$f_x(x) = \frac{(\alpha^2-\beta^2)^{\frac{\lambda}{2}} K_{\lambda-\frac{1}{2}}(\alpha\sqrt{\delta^2+(x-\mu)^2}\exp(\beta(x-\mu))}{\sqrt{2\pi}\alpha^{\lambda-\frac{1}{2}}\delta^\lambda K_\lambda(\delta\sqrt{\alpha^2-\beta^2})(\sqrt{\delta^2-(x-\mu)^2})^{\frac{1}{2}-\lambda}} \tag{1}$$

In the above expression, $K_j$ denotes the modified Bessel function of the third kind of order $j$ (Abramowitz and Stegun, 1972).The relation between the parameters should satisfied that: when $\lambda > 0$, then $\delta \geq 0$, $|\beta| < \alpha$; when $\lambda = 0$, then $\delta > 0$, $|\beta| < \alpha$; when $\lambda < 0$, then $\delta > 0$, $|\beta| \leq \alpha$.

Barndorff-Nielsen and Blæsild (1981) also stated that the GH distribution can be represented as a mixed distribution of the normal variance-mean and the generalized inverse Gaussian, which means that the generalized hyperbolic variable X can be expressed as:

$$X = \mu + \beta Z + \sqrt{Z}Y, \tag{2}$$

where $Y \sim N(0,1)$, $Z \sim GIG(\lambda, \delta, \gamma)$, with $Y$ and $Z$ independent and $\gamma = \sqrt{\alpha^2 - \beta^2}$. Equation (2) infers that $X|Z = z \sim N(\mu + \beta z, z)$. $GIG(\cdot)$ denotes the Generalized Inverse Gaussian (GIG) proposed by Barndorff-Nielsen (1977), having density:

$$f(z;\lambda,\delta,\gamma) = \left(\frac{\gamma}{\delta}\right)^\lambda \frac{z^{\lambda-1}}{2K_\lambda(\gamma\delta)} exp\left\{-\frac{1}{2}(\delta^2 z^{-1} + \gamma^2 z)\right\} \tag{3}$$

Letting $\lambda = -\frac{1}{2}$, then GH distribution evolves into a special case, namely the normal inverse Gaussian distribution. The modified Bessel function has properties (Blæsild, 1981): $K_{\frac{1}{2}}(x) = 2^{-\frac{1}{2}}\sqrt{\pi}x^{-\frac{1}{2}}\exp(-x)$ and $K_v(x) = K_{-v}(x)$. Then the density function of NIG will be:

$$f_x(x) = \frac{\delta\alpha \exp(\delta\sqrt{\alpha^2-\beta^2})K_1(\alpha\sqrt{\delta^2+(x-\mu)^2}\exp(\beta(x-\mu))}{\pi\sqrt{\delta^2+(x-\mu)^2}}, \delta > 0, 0 < |\beta| < \alpha \tag{4}$$

If $X$ obeys the NIG distribution, it can be written as $X \sim NIG(\mu, \delta, \alpha, \beta)$, where $\mu, \delta, \alpha, \beta$ respectively represent the parameters of location, scale, tail, and skewness that determine the form of distribution. The attractive property of the NIG distribution is that under the convolution of independent random variables $Y_1$ and $Y_2$, the NIG distribution is closed:

$$Y_1 \sim NIG(\mu_1, \delta_1, \alpha, \beta), Y_2 \sim NIG(\mu_2, \delta_2, \alpha, \beta) \Rightarrow Y_1 + Y_2 \sim NIG(\mu_1 + \mu_2, \delta_1 + \delta_2, \alpha, \beta) \quad (5)$$

When there exists a variable $Y = \sum_{i=1}^{n} Y_i$, and $Y_i, i = 1, \cdots, n$ is independent and NIG-distributed, the density function of $Y$ then can be:

$$g_Y(y) = \frac{n\delta\alpha\exp(n\delta\sqrt{\alpha^2-\beta^2})K_1(\alpha\sqrt{n^2\delta^2+(y-n\mu)^2}\exp(\beta(y-n\mu))}{\pi\sqrt{n^2\delta^2+(y-n\mu)^2}}, \delta > 0, 0 < |\beta| < \alpha \quad (6)$$

Let $R_{\tau,t}$ denote the $\tau^{th}$ observation of intraday return on the $t$th day, where $t = 1, \ldots, T$, $\tau = 1, \ldots, N$, and N varies due to the frequency of the intraday return. If it is assumed that the high-frequency returns on day $t$ are independent of each other, and are all NIG-distributed with the same parameters $\alpha$ and $\beta$, where $\alpha$ and $\beta$ are determined by the NIG distribution of daily return on day $t$, then the distribution of this daily return could be re-estimated by combining the intraday return, since $R_t = \sum_{\tau=1}^{N} R_{\tau,t}$. This practice is regarded as a possible way to adjust or improve the fitting of return, and the semi-additivity of NIG makes it much easier to operate.

**2.2 NIG-DCS Model**

We establish a model driven by DCS to estimate the time-varying parameters in equation (4). The score-driven mechanism is widely adopted because the score naturally drive the parameters to update by linking their dynamics to the likelihood probability of the historical observations (Creal et al., 2012). Blasques, et al. (2015) also stated that DCS method may be the optimal method in information theory since its updating mechanism could significantly reduce the differences between the estimated conditional density and the actual conditional density. Although GARCH models has been proved to be effective in fitting financial returns and giving accurate forecasts in previous studies (Creal, Koopam and Lucas, 2011; Gao and Zhou, 2016; Avdulaj and Barunik, 2015), Harvey and Sucarrat (2014) believed that the DCS model can be superior to it because DCS works better on estimating the heavy tails of returns. Additionally, the scheme of DCS allows for the one-step prediction that could form a series out-of-sample VaR forecasts by apply a rolling-window procedure. In DCS, the common form of return (Creal et al., 2011, 2013) is:

$$R_t = \mu_t + v_t = \mu_t + \exp(\lambda_t)\epsilon_t, \quad (7)$$

where $\mu_t$, $\exp(\lambda_t) = \delta_t$ respectively represent the location and scale of the distribution, and the exponential form of $\lambda_t$ makes the scale always larger than 0. The main object of DCS modeling is the random volatility $\epsilon_t$. If $R_t \sim NIG(\mu_t, \delta_t, \alpha_t, \beta_t)$, in order to satisfy $\delta_t > 0$ as well as $0 < |\beta_t| < \alpha_t$, let:

$$\delta_t = e^{\lambda_t}, \alpha_t = e^{v_t - \lambda_t}, \beta_t = e^{v_t - \lambda_t} \cdot \tanh(\eta_t) = e^{v_t - \lambda_t} \cdot \frac{e^{\eta_t} - e^{-\eta_t}}{e^{\eta_t} + e^{-\eta_t}} \tag{8}$$

Namely, $\epsilon_t \sim NIG(0,1, e^{v_t}, e^{v_t} \tanh(\eta_t))$. From equation (4), the logarithmic conditional probability density of $R_t$ becomes:

$$\ln f(r_t | r_1, \ldots, r_{t-1}) = \delta_t \sqrt{\alpha_t^2 - \beta_t^2} + \ln \alpha_t + \ln \delta_t - \ln \pi - \frac{1}{2} \ln(\delta_t^2 + (r_t - \mu_t)^2) +$$

$$\ln K_1 \left( \alpha_t \sqrt{\delta_t^2 + (r_t - \mu_t)^2} \right) + \beta_t (r_t - \mu_t) = e^{v_t} \sqrt{1 - \tanh^2(\eta_t)} + v_t - \ln \pi - \frac{1}{2} \ln(e^{2\lambda_t} + (r_t - \mu_t)^2) +$$

$$\ln K_1 \left( e^{v_t - \lambda_t} \sqrt{e^{2\lambda_t} + (r_t - \mu_t)^2} \right) + e^{v_t - \lambda_t} \tanh(\eta_t)(r_t - \mu_t) \tag{9}$$

For the four time-varying parameters in (9), the law of motion is specified in terms of DCS framework. The main feature of DCS is that the evolution of the time-varying vector is driven by the score of the conditional distribution (9), an autoregressive component, and other possible explanatory variables. The score refers to the first-order partial derivative of the conditional distribution with respect to the time-varying parameter. Since there may exist annual cycle structure in returns during a long period of time, in order to ensure the independence of the time series, we refer to the practice of Harvey and Ito (2020) and consider adding the seasonal factor $q_t$ to the autoregressive equation of scale parameter. The deterministic $q_t$ can be easily obtained by decomposing the time series. Then the dynamic laws of the four time-varying parameters $\mu_t$, $\lambda_t$, $v_t$, and $\eta_t$ are respectively specified as:

$$\begin{cases} \mu_t = A_1 + B_1 \mu_{t-1} + C_1 s_{\mu_t} \\ \lambda_t = A_2 + B_2 \lambda_{t-1} + C_2 s_{\lambda_t} + q_t \\ v_t = A_3 + B_3 v_{t-1} + C_3 s_{v_t} \\ \eta_t = A_4 + B_4 \eta_{t-1} + C_4 s_{\eta_t} \end{cases}, \tag{10}$$

where $s_{\mu_t}$, $s_{\lambda_t}$, $s_{v_t}$ and $s_{\eta_t}$ are the so-called score factors of the corresponding parameters and they can be calculated by the following formulas:

$$\frac{\partial \ln[f_t(r_t | r_1, \ldots, r_{t-1})]}{\partial \mu_t} = -e^{v_t - \lambda_t} \tanh(\eta_t) + \frac{(r_t - \mu_t) e^{-\lambda_t}}{e^{\lambda_t} + (r_t - \mu_t)^2 e^{-\lambda_t}} + \frac{(r_t - \mu_t) e^{v_t - 2\lambda_t}}{\sqrt{1 + (r_t - \mu_t)^2 e^{-2\lambda_t}}} \cdot$$

$$\frac{K^{(0)} \left[ e^{v_t - \lambda_t} \sqrt{1 + (r_t - \mu_t)^2} \right] + K^{(2)} \left[ e^{v_t - \lambda_t} \sqrt{1 + (y_t - \mu_t)^2} \right]}{2K^{(1)} \left[ e^{v_t - \lambda_t} \sqrt{1 + (y_t - \mu_t)^2} \right]} \tag{11}$$

$$s_{\mu_t} = \frac{\partial \ln[f_t(r_t | r_1, \ldots, r_{t-1})]}{\partial \mu_t} \cdot e^{2\lambda_t} \tag{12}$$

$$s_{\lambda_t} = \frac{\partial \ln[f_t(r_t|r_1,...,r_{t-1})]}{\partial \lambda_t} = -1 + \frac{(r_t-\mu_t)^2 e^{-2\lambda_t}}{\sqrt{1+(r_t-\mu_t)^2 e^{-2\lambda_t}}} -$$

$$e^{v_t-\lambda_t}\tanh(\eta_t)(r_t - \mu_{\tau,t}) + \frac{(r_t-\mu_t)^2 e^{v_t-2\lambda_t}}{\sqrt{1+(r_t-\mu_t)^2 e^{-2\lambda_t}}} \cdot \frac{K^{(0)}\left[e^{v_t-\lambda_t}\sqrt{1+(r_t-\mu_{\tau,t})^2}\right]+K^{(2)}\left[e^{v_t-\lambda_t}\sqrt{1+(r_t-\mu_t)^2}\right]}{2K^{(1)}\left[e^{v_t-\lambda_t}\sqrt{1+(r_t-\mu_t)^2}\right]} \quad (13)$$

$$s_{v_t} = \frac{\partial \ln[f_t(r_t|r_1,...,r_{t-1})]}{\partial v_t} = 1 + e^{v_t}\sqrt{1-\tanh^2(\eta_t)} - e^{v_t-\lambda_t}\tanh(\eta_t)(r_t - \mu_t) -$$

$$e^{v_t}\sqrt{1+(r_t-\mu_t)^2 e^{-2\lambda_t}} \cdot \frac{K^{(0)}\left[e^{v_t-\lambda_t}\sqrt{1+(y_t-\mu_t)^2}\right]+K^{(2)}\left[e^{v_t-\lambda_t}\sqrt{1+(r_t-\mu_t)^2}\right]}{2K^{(1)}\left[e^{v_t-\lambda_t}\sqrt{1+(r_t-\mu_t)^2}\right]} \quad (14)$$

$$s_{\eta_t} = e^{v_t-\lambda_t}\text{sech}^2(\eta_t) \cdot (r_t - \mu_t) - e^{v_t}\tanh(\eta_t)\text{sech}(\eta_t) \quad (15)$$

The steps of modeling intraday return using NIG-DCS is similar to the above. But as mentioned in 2.1, if it is assumed that the intraday returns on day $t$ are NIG-distributed with the same tail parameters and skewness parameters that determined by the daily return on day $t$, namely, $R_{\tau,t} \sim NIG(\mu_{\tau,t}, e^{\lambda_{\tau,t}}, e^{v_{\tau,t}-\lambda_{\tau,t}} = e^{v_t-\lambda_t}, e^{v_{\tau,t}-\lambda_{\tau,t}}\tanh(\eta_{\tau,t}) = e^{v_t-\lambda_t} \cdot \tanh(\eta_t))$, then $\eta_{\tau,t}$ could be directly determined by $\eta_t$, while $\mu_{\tau,t}$ and $v_{\tau,t}$ are still driven by DCS:

$$\begin{cases} \mu_{\tau,t} = A_5 + B_5\mu_{\tau,t-1} + C_5 s_{\mu_{\tau,t}} \\ v_{\tau,t} = A_6 + B_6 v_{\tau,t-1} + C_6 s_{v_{\tau,t}} \end{cases}, \quad (16)$$

then $\lambda_{\tau,t}$ can be indirectly obtained from $e^{v_t-\lambda_t}$ and $v_{\tau,t}$. The variation of $\lambda_{\tau,t}$ ensures that intra-day returns have constant tail and skewness parameters when $v_{\tau,t}$ changes. Here, the reason why we let $v_{\tau,t}$ rather than $\lambda_{\tau,t}$ be DCS-driven is that the estimated intraday return under this scheme is much closer to the actual return. This setting is proved to be more reasonable by empirical results. Finally, maximum likelihood method could hence be performed to estimate the parameter $\theta$ involving in the NIG-DCS model. Formally:

$$\hat{\theta} = argmax \sum_{t=1}^{n} \ln[f_t(r_t|r_1,...,r_{t-1})], \quad t = 1,2,3...T \quad (17)$$

### 2.3 Intraday-return-based Estimation for Daily VaR

Predicting daily VaR in light of intraday return requires careful consideration due to the non-normality and randomness of HF data. Since high-frequency data is often affected by market microstructure noise (Andersen et al., 2001a), which sampling frequency is used to obtain intraday returns is also a key issue. Two methods of estimating the distribution of daily return incorporating intraday return are introduced in the following.

From the assumption of intraday return and the special property of NIG mentioned in section 2.1, it follows that $R_t \sim NIG\left(\sum_{\tau=1}^{N} \mu_{\tau,t}, \sum_{\tau=1}^{N} e^{\lambda_{\tau,t}}, e^{v_t - \lambda_t}, e^{v_t - \lambda_t} \cdot \tanh(\eta_t)\right)$, since $R_t = \sum_{\tau=1}^{N} R_{\tau,t}$. This means that the distribution of daily returns is adjusted or improved through the changes in location and scale parameters. Common statistical software such as R, etc. can explicitly calculate the quantile of the distribution based on the known specification and known parameters. Since VaR at the level α is essentially the $\alpha^{th}$ quantile of the distribution, the daily VaR can hence be obtained by this method.

The other method is non-parametric and is based on bootstrap, which can generate a distribution of daily returns by repeated random sampling. Compared with the parametric model, this method only assumes that the intraday returns follow the NIG distribution. But there is no need to assume that its tail parameters and skewness parameters are consistent with those in the distribution of the corresponding daily return. And similarly, the deterministic component of seasonal factor is also included in the model to reduce the possible correlation of HF returns. Hence, equation (16) in this situation should be expanded as:

$$\begin{cases} \mu_{\tau,t} = A_5 + B_5 \mu_{\tau,t-1} + C_5 s_{\mu_{\tau,t}} \\ \lambda_{\tau,t} = A_6 + B_6 \lambda_{\tau,t-1} + C_6 s_{\lambda_{\tau,t}} + q_{\tau,t} \\ v_{\tau,t} = A_6 + B_6 v_{\tau,t-1} + C_6 s_{v_{\tau,t}} \\ \eta_{\tau,t} = A_4 + B_4 \eta_{\tau,t-1} + C_4 s_{\eta_{\tau,t}} \end{cases} \quad (18)$$

The specific steps of bootstrap can see from Song et al.(2021).

Both of the above methods can fit daily returns combining the information of intraday information. But when conducting empirical study, we consider the parametric method to highlight the attractive property of NIG and enhance the efficiency of estimation.

The proposed NIG-DCS-VaR model is a tool to measure daily VaR based on intraday information. The model first uses NIG distribution to fit intraday returns and daily returns, which can well capture the tail risk of returns; then, in light of the mechanism of the dynamic conditional score, the motion laws of time-varying parameters are specified, and the maximum likelihood method is applied to estimate the optimal results; after that, we propose two methods: parametric method and bootstrap method, which can integrate the estimated results of multiple intraday return to fit the daily return. Finally, daily VaR is easily obtained by calculating the quantile of the fitted distribution.

Due to the adoption of the DCS mechanism, the NIG model is more sensitive to the volatility of risk, and the incorporation of high-frequency information allows it to measure risk from a more microscopic perspective. Therefore, the NIG-DCS-VaR model can theoretically provide a more accurate forecast of daily VaR, thereby contributing to the improvement of VaR measurement tools.

**2.4 Backtesting and MCS for VaR**

In order to verify whether the results of VaR estimates are consistent and reliable, several appropriate backtestings are mainly used. We will apply the LR of Unconditional Coverage test (LRUC), the LR of conditional Coverage test (LRCC), and the Dynamic Quantile test (DQ) these three backtestings to measure the dynamic VaR forecasts obtained by NIG-DCS-VaR model. Kupiec (1995) proposed LRUC as a statistical tool to determine whether the model will be accepted or rejected by judging whether the number of VaR exceptions, i.e. days when negative returns exceed VaR estimates, is reasonable or not. Christoffersen (1998) came up with the LRCC to test not only the frequency of VaR violations but also the time when they occur. DQ test proposed by Engle and Manganelli (2004) is mainly used to check the independence of the VaR series. It constructs indicators of occurrence of the extreme events. If there exists correlation between these indicators, the dynamic VaR model will be proved to be invalid.

However, it is difficult to intuitively compare the quality of the models only relying on backtestings. Hansen et al. (2011) developed the Model Confidence Set (MCS) procedure to compare the performance of a given set of VaR series. The Hansen's procedure works like a more comprehensive evaluation tool so we could perform MCS to evaluate the accuracy and effectiveness of several VaR models.

**3. Data and data processing**

**3.1 Construction of Return Series**

We choose the Shanghai SE Composite Index (SH000001) and the Shenzhen SE Component Index (SZ399001) in the Chinese stock market as the empirical objects in application due to their position of vane of the market. We select the daily closing prices from January 5, 2009 to December 31, 2015 that includes 1700 trading days to obtain daily returns. We also obtain the 1-minute intraday price data and thus form the HF dataset of 10-minute return, 20-minute return, 30-minute return and 40-minute return through different sampling frequency. And the size of HF dataset are 1700×24, 1700×12, 1700×8 and 1700×6, respectively.

All data can be obtained from WIND. The daily return is defined as $R_t = logP_t - logP_t$, and the intraday return can be thus obtained by $R_{\tau,t} = logP_{\tau,t} - logP_{\tau-1,t}$, where $P$ refers to the closing price at that time.

In this empirical work, we will first reverse the sign of historical returns and calculate the VaR for the extreme value along the right tail, in order to display the great losses more intuitively. Table 1 shows the descriptive statistical results of the historical daily return of SH000001 and SZ399001 during January 5, 2009 to December 31, 2015, and the unit root test shows that they are all first-order stationary, thereby ensuring the stability of the return series.

Table 1.Descriptive Statistics, Daily Returns, SH000001 and SZ399001, 2009–2015.

|  | sh000001 $p_t$ | sh000001 $\ln(p_t/p_{t-1})$ | sz399001 $p_t$ | sz399001 $\ln(p_t/p_{t-1})$ |
|---|---|---|---|---|
| Start date | 2009-01-05 | 2009-01-05 | 2009-01-05 | 2009-01-05 |
| End date | 2015-12-31 | 2015-12-31 | 2015-12-31 | 2015-12-31 |
| Sample size | 1700 | 1700 | 1700 | 1700 |
| Minimum | 1880.72 | -0.0815 | 6634.88 | -0.1103 |
| Maximum | 5178.19 | 0.0547 | 18211.76 | 0.05701 |
| Average | 2685.763 | -1e-04 | 10502.45 | -1e-04 |
| Standard deviation | 603.0049 | 0.0128 | 2160.589 | 0.0151 |
| Skewness | 1.3288 | -0.7813 | 0.4623 | -0.8916 |
| kurtosis | 2.0434 | 4.6337 | -0.1824 | 5.4299 |
| ADF statistic | -1.783(0.6703) | -8.2203(<0.01**) | -2.4815(0.3745) | -11.595(<0.01**) |

Note: *, **, and *** represent statistical significance levels of 5%, 1%, and .1%, respectively. The p-value of the relevant test is indicated by the value in parentheses.

In section 2.3, we introduce two methods of integrating the information of intraday returns to form daily returns in details. But it is worth noting that the parametric method requires the return series to be independent of each other. Therefore, Pearson's correlation test is applied to check whether the data used in the empirical study meets this requirement. Table 2 lists the p-value of the test between two adjacent series of intraday return.

The proportion of sequence pairs that do not meet the independence requirement under each frequency is also listed in the last row in Table 2. The results tell that there exists some correlation between the adjacent series of intraday returns. The dependence of 20min-HF is particularly significant, but that of 30min-HF is relatively weakest. Although it is reasonable to assume that the intraday returns are mutually independent according to the efficient market hypothesis, we add the seasonal factor to the dynamic equation of the scale parameter, which could reduce the correlation between high-frequency series to a certain extent.

Table 2. Results of Pearson's correlation test for intraday return, SH000001 and SZ399001, 2009–2015.

| Pair of returns | SH000001 | | | | SZ399001 | | | |
|---|---|---|---|---|---|---|---|---|
| | 10min-HF | 20min-HF | 30min-HF | 40min-HF | 10min-HF | 20min-HF | 30min-HF | 40min-HF |
| 1-2   | 0(***)   | 0(***)   | 0.89    | 0.98    | 0(***)   | 0(***)   | 0.52    | 0.78    |
| 2-3   | 0(***)   | 0(***)   | 0(***)  | 0.34    | 0(***)   | 0(***)   | 0.14    | 0.14    |
| 3-4   | 0(***)   | 0(***)   | 0.27    | 0(***)  | 0(***)   | 0.3      | 0(***)  | 0(***)  |
| 4-5   | 0(***)   | 0.57     | 0(***)  | 0(***)  | 0(***)   | 0(***)   | 0(***)  | 0(***)  |
| 5-6   | 0.22     | 0(***)   | 0.55    | 0(***)  | 0.46     | 0(***)   | 0.85    | 0(***)  |
| 6-7   | 0.01(**) | 0(***)   | 0(***)  |         | 0.04(**) | 0(***)   | 0(***)  |         |
| 7-8   | 0.03(**) | 0(***)   | 0(***)  |         | 0.32     | 0.28     | 0(***)  |         |
| 8-9   | 0.58     | 0(***)   |         |         | 0.02(**) | 0.01(**) |         |         |
| 9-10  | 0.88     | 0.36     |         |         | 0.54     | 0.34     |         |         |
| 10-11 | 0(***)   | 0(***)   |         |         | 0(***)   | 0(***)   |         |         |
| 11-12 | 0(***)   | 0(***)   |         |         | 0(***)   | 0(***)   |         |         |
| 12-13 | 0(***)   |          |         |         | 0(***)   |          |         |         |
| 13-14 | 0(***)   |          |         |         | 0(***)   |          |         |         |
| 14-15 | 0(***)   |          |         |         | 0(***)   |          |         |         |
| 15-16 | 0.85     |          |         |         | 0.58     |          |         |         |
| 16-17 | 0.07     |          |         |         | 0.1      |          |         |         |
| 17-18 | 0(***)   |          |         |         | 0(***)   |          |         |         |
| 18-19 | 0.28     |          |         |         | 0.14     |          |         |         |
| 19-20 | 0.35     |          |         |         | 0.77     |          |         |         |
| 20-21 | 0.4      |          |         |         | 0.65     |          |         |         |
| 21-22 | 0.29     |          |         |         | 0.78     |          |         |         |
| 22-23 | 0(***)   |          |         |         | 0(***)   |          |         |         |
| 23-24 | 0.01     |          |         |         | 0        |          |         |         |
| ratio | 60.87%   | 81.82%   | 57.14%  | 60.00%  | 60.87%   | 72.73%   | 57.14%  | 60.00%  |

Note: ∗, ∗∗, and ∗∗∗ represent statistical significance levels of 5%, 1%, and .1%, respectively. All the values of the table above indicate the p-values of the Pearson's correlation test, which determines whether reject the null hypothesis that the correlation between two series is equal to 0.

As explained in 2.2, following the practice of Harvey and Ito (2013), we add the periodic factor to the score-driven equation of the scale parameter $\lambda$, to eliminate the influence of the periodic structure on the return volatility. Since we assume that the tail parameter as well as the skewness parameter is consistent with the daily return, $\lambda$ of intraday return will be determined by that of daily return rather than its own DCS law. Therefore, the deterministic component in daily return will help to indirectly identify the deterministic component in intraday returns. And this so-called deterministic component may be the possible factor that makes the intraday series be correlated with each other. In other word, considering the periodic component as an extra explanatory variable that influence the scale parameter, could not only promote the accuracy of fitting, but also eliminate the impact of correlation between series of high-frequency return to a certain extent, and thus ensure the feasibility of the parametric method.

**3.2 Vacation And Weekend Effect**

Generally, the "vacation and weekend effect" refers to the phenomenon that the opening price after the end of a long closed period may be remarkably different from the closing price before the start of the closed period. This abnormal fluctuation may lead to "fake" extreme returns that can hardly represent the actual information of the market. So there is a need to adjust returns affected by this kind of effect before constructing models. The practice of this adjustment is the same as what mentioned in Song et al.(2021)'s work.

**3.3 Overnight Effect**

The "overnight effect" refers to a phenomenon that due to the accumulation of night information, a significantly different change in stock returns will occur after the stock market opens the next day than in other periods. French and Roll (1986) believed that the continuous accumulation of private information during the non-trading period leads to cognitive biases of traders and thus active transactions during the trading period. Greene and Watts (1996) proved that not only private information, but public information will also cause investors to trade more actively, resulting in greater volatility in opening moment of the market. In our case, we believe that there is no need to eliminate this "overnight effect" because it contains the original information of price change which should be considered in risk management.

**3.4 Periodic Structure**

Daily returns will be more or less affected by the seasons. We assume that the period of this impact is 366 days (including the special case of leap years). Since the returns in the empirical study are logarithmic, considering the missing values on holidays and February 29 in non-leap years to be zero is reasonable, which means no price change in those days. Based on the historical data, we use moving average method to decompose the trend item, and then averages the corresponding values at each day in the cycle to obtain the deterministic annual periodic structure. It is also feasible to directly average the corresponding returns of each unit day to obtain the cycle component, and the result is similar to the previous method. Figure 1 shows the yearly periodic structure of SH000001 and SZ399001.

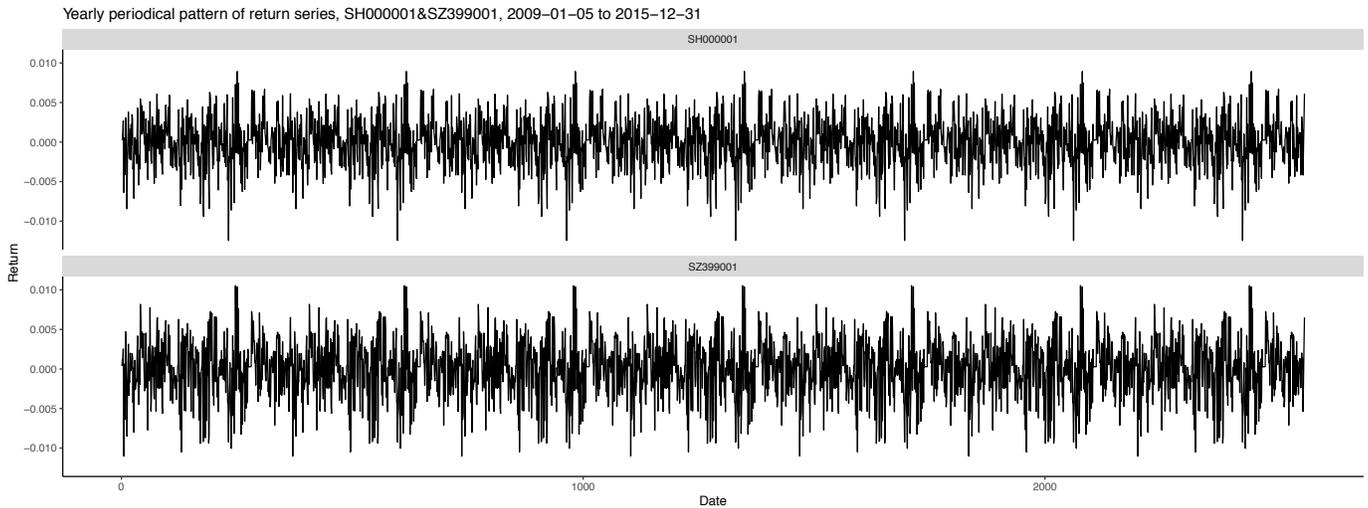

Figure 1. Seasonal cycle for return, SH000001&SZ399001.

## 4. Empirical Results

### 4.1 In-sample VaR Forecasts

The data in the window from January 5, 2009 to December 31, 2014 is used to build the model and obtain the in-sample VaR forecasts. The in-sample results will imply the validity of daily return integrated by intraday return with different frequency. This will also help to identify the relatively better choice of HF data in our application.

For the in-sample VaR generated by the models based on different frequencies of HF, we apply several classic backtestings to compare their performance at $\alpha = \{0.90, 0.91, 0.92, ..., 0.97, 0.98, 0.99\}$. The results of SH000001 and SZ399001 are listed in Table 4 and Table 5, respectively (To keep the list concise, we only give the four cases where $\alpha$ is 0.93, 0.95, 0.97 and 0.99 in the table, see the appendix A for the complete list). For SH000001, taking the statistics and the p-values at different level of α into account, it can be judged that the NIG-DCS-VaR based on daily return (LF) is basically weaker than HF-based VaR (except 10 min HF-based VaR) in coverage ability and independence. MCS rankings also prove the superiority of HF-based models. This means that incorporating the information of intraday returns into the model can improve the accuracy of VaR forecasts to a certain extent, but if the used intraday returns are with a very high frequency, the distribution of daily returns may be overestimated and hence the VaR forecast could be overestimated. Furthermore, when $\alpha \leq 0.95$, 40min-based VaR perform relatively better in LRUC test, LRCC test, DQ test, and MCS test than other HF-based VaR, followed by 30min-based VaR; but when $\alpha > 0.95$, 30min-based VaR forecasts perform relatively best.

Table 3. Back testing results of in-sample daily VaR estimates, SH000001.

| Model | Alpha | LRUC statistics | LRCC statistics | DQ statistics | MCS rank ($\alpha = 0.15$) | Alpha | LRUC statistics | LRCC statistics | DQ statistics | MCS rank ($\alpha = 0.15$) |
|---|---|---|---|---|---|---|---|---|---|---|
| VaR-day | 0.93 | 2.82 (0.09) | 4.2 (0.12) | 5.55 (0.59) | 4 | 0.97 | 1.17 (0.28) | 1.41 (0.49) | 2.44 (0.93) | 5 |
| VaR-40minhq | | 0.48 (0.49) | 2.79 (0.25) | 5.9 (0.55) | 1 | | 0.5 (0.48) | 0.83 (0.66) | 1.88 (0.97) | 2 |
| VaR-30minhq | | 2.03 (0.15) | 3.62 (0.16) | 4.8 (0.68) | 2 | | 0.5 (0.48) | 0.83 (0.66) | 1.88 (0.97) | 1 |
| VaR-20minhq | | 3.76 (0.05) | 4.94 (0.08) | 5.6 (0.59) | 3 | | 0.12 (0.73) | 0.56 (0.76) | 4.06 (0.77) | 3 |
| VaR-10minhq | | 0.48 (0.49) | 1.21 (0.55) | 11.71 (0.11) | 5 | | 0.43 (0.51) | 1.12 (0.57) | 5.73 (0.57) | 4 |
| | | | | | (0.66) | | | | | (0.56) |
| VaR-day | 0.95 | 1.23 (0.27) | 2.07 (0.35) | 4.14 (0.76) | 4 | 0.99 | 0.3 (0.58) | 0.41 (0.81) | 0.69 (1) | 4 |
| VaR-40minhq | | 0.29 (0.59) | 1.48 (0.48) | 3.3 (0.86) | 1 | | 0.3 (0.58) | 0.41 (0.81) | 0.64 (1) | 2 |
| VaR-30minhq | | 0.68 (0.41) | 1.68 (0.43) | 3.21 (0.87) | 2 | | 0.3 (0.58) | 0.41 (0.81) | 0.64 (1) | 1 |
| VaR-20minhq | | 1.23 (0.27) | 2.07 (0.35) | 3.72 (0.81) | 3 | | 0.32 (0.58) | 0.41 (0.81) | 0.64 (1) | 3 |
| VaR-10minhq | | 0.07 (0.79) | 2.13 (0.34) | **15.86 (0.03\*)** | 5 | | 1.12 (0.29) | 1.29 (0.52) | 3.7 (0.81) | 5 |
| | | | | | (0.33) | | | | | (0.35) |

Note: ∗, ∗∗, and ∗∗∗ represent statistical significance levels of 5%, 1%, and .1%, respectively. The rank tells the superiority of these four models under a default level of $\alpha$ and p-value helps to prove the non-rejection of this superiority. The p-value of the relevant test is indicated by the value in parentheses. Bold text indicates rejections at the * probability level.

For SZ399001, no matter at which level, the 30min-HF-based VaR forecasts perform significantly better than others in terms of two tests for coverage ability, but their independence show a disadvantage. However, MCS rankings give the comprehensive results of measurement and still support 30min-HF-based model. On the whole, the 30min-based-NIG-DCS model is more stable and can maintain the accuracy of VaR forecasts at a high risk level. Table 2 also proves that the intraday return at 30min intervals has the weakest correlation. Therefore, we believe that 30min-intraday returns can be considered as the relatively best modeling object. We will construct NIG-DCS model based on 30min-intraday returns to obtain out-of-sample forecasts and compare them with those from other benchmark models.

Table 4. Back testing results of in-sample daily VaR estimates, SZ399001.

| Model | Alpha | LRUC statistics | LRCC statistics | DQ statistics | MCS rank ($\alpha = 0.15$) | Alpha | LRUC statistics | LRCC statistics | DQ statistics | MCS rank ($\alpha = 0.15$) |
|---|---|---|---|---|---|---|---|---|---|---|
| VaR-day | 0.93 | 5.25 (0.02*) | 5.34 (0.07) | 5.98 (0.54) | 3 | 0.97 | 6.09 (0.01*) | 6.09 (0.05) | 2.97 (0.89) | 4 |
| VaR-40minhq | | 8.49 (0.01**) | 8.51 (0.01*) | 5.97 (0.54) | 4 | | 1.84 (0.17) | 1.86 (0.39) | 3.6 (0.82) | 3 |
| VaR-30minhq | | 0 (1) | 0.48 (0.79) | 15.18 (0.03*) | 1 | | 1.15 (0.28) | 2.66 (0.26) | 11.85 (0.1) | 1 |
| VaR-20minhq | | 14.51 (0.01**) | 14.51 (0.01**) | 7.23 (0.41) | 5 | | 6.09 (0.01*) | 6.09 (0.05) | 2.97 (0.89) | 5 |
| VaR-10minhq | | 3.09 (0.08) | 3.27 (0.19) | 14.93 (0.04*) | 2 | | 0 (1) | 0.19 (0.91) | 25.58 (<0.01**) | 2 |
| | | | | | (0.55) | | | | | (0.38) |
| VaR-day | 0.95 | 10.26 (0.01**) | 10.26 (0.01*) | 5.05 (0.65) | 4 | 0.99 | 2.01 (0.16) | 2.01 (0.37) | 0.97 (1) | 4 |
| VaR-40minhq | | 4.95 (0.03*) | 4.97 (0.08) | 4.33 (0.74) | 3 | | 2.01 (0.16) | 2.01 (0.37) | 0.97 (1) | 3 |
| VaR-30minhq | | 0 (1) | 1.51 (0.47) | 20.37 (<0.01**) | 1 | | 0 (1) | 0.02 (0.99) | 12.81 (0.06) | 1 |
| VaR-20minhq | | 10.26 (0.01**) | 10.26 (0.01*) | 5.05 (0.65) | 5 | | 2.01 (0.16) | 2.01 (0.37) | 0.97 (1) | 5 |
| VaR-10minhq | | 0.98 (0.32) | 1.16 (0.56) | 17.26 (0.02*) | 2 | | 0.78 (0.38) | 0.87 (0.65) | 49.64 (<0.01**) | 2 |
| | | | | | (0.44) | | | | | (0.45) |

Note: ∗, ∗∗, and ∗∗∗ represent statistical significance levels of 5%, 1%, and .1%, respectively. The rank tells the superiority of these four models under a default level of $\alpha$ and p-value helps to prove the non-rejection of this superiority. The p-value of the relevant test is indicated by the value in parentheses. Bold text indicates that it fails the test.

Figure 2 compares the in-sample VaR estimates between the daily-return-based and 30min-HF-based. As shown in the figure, 30min-based model has an obvious smaller overall forecast error, and its coverage ability is much more in line with expectations, making it perform well in the LRCC and LRUC tests. We originally thought that the daily VaR estimated by integrating intraday information will be more sensitive to risks and should reflect greater fluctuation, but the results indicates that the addition of high-frequency information reduces the volatility of VaR in the face of extreme returns, which is out of our expectation. Compared with the results based on LF, HF-based NIG-DCS-VaR behave more smoothly.

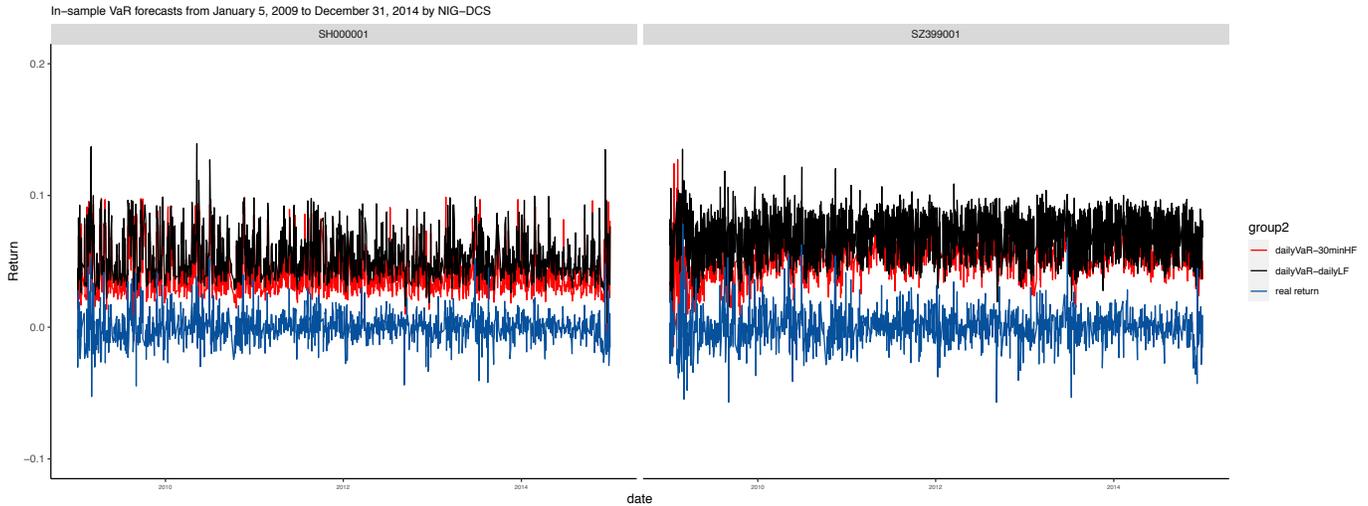

Figure 2. Comparison between HF-based NIG-DCS-VaR and LF-based NIG-DCS-VaR, SH000001 & SZ399001.

An effective DCS model requires that the score series of involved time-varying parameters should not have autocorrelation. Lagrange multiplier (LM) test proposed by Harvey and Thiele (2016) are usually used to check the autocorrelation of time series and we also apply it to check the effectiveness of NIG-DCS models we construct. The p-values of the LM test for the in-sample scores are listed in Appendix A, including both daily-return-based results and 30min-return-based results. All the p-values are greater than 0.05, indicating that there is no significant autocorrelation in the in-sample scores of different parameters that are under different DCS models. Figure 3 briefly shows the correlograms of scores for the four parameters estimated based on the daily return, as well as the correlograms of scores for two DCS-driven parameters of the first 30min returns and the last 30min returns.

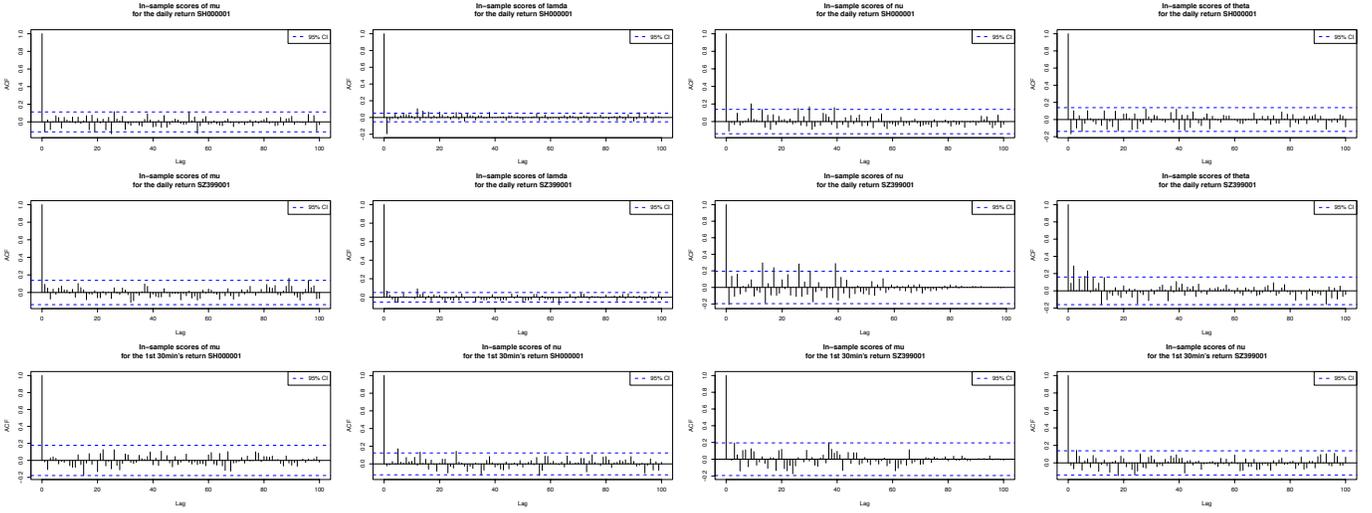

Figure 3. Correlograms of in-sample scores for fitted daily return and 1st 30min's return, SH000001&SZ399001. The first two rows represent score correlograms of the four parameters of daily return, and the bottom represents the score correlograms of the two parameters of the first 30min's return.

**4.2 Out-of-sample VaR Forecasts**

From May to August in 2015, China's stock market experienced a "disaster" since two rounds of cliff-shaped declines occurred during a very short period. We believe that make the out-of-sample VaR forecasts in this time window has great empirical significance. We expect to utilize the NIG-DCS model based on the intraday data with 30min-frequency, to apply a rolling-window scheme to obtain a time series of VaR forecasts at different confidence levels, i.e. $\alpha = \{0.90, 0.91, 0.92, 0.93, 0.94, 0.95, 0.96, 0.97, 0.98, 0.99\}$. Let $n$ denote the length of VaR to be predicted, $m$ denote the size of the available sample and $s$ denote the length of the rolling window. Then we have the sequence of forecasts $\{VaR_t^\alpha, t = s + 1, s + 2 ..., s + n\}$, where each prediction is obtained considering the observations that incorporate the intraday return, $\{R_{\tau,t-s}\}_{\tau=1}^8, \{R_{\tau,t-s+1}\}_{\tau=1}^8 ..., \{R_{\tau,t-1}\}_{\tau=1}^8$. We produce $n = 244$ daily VaR forecasts for January 5 to December 31 in 2015 by considering the size of the window to be $s = 1456$, and the initial one starts from January 5, 2009 and ends with December 31, 2014. Figure 4 presents the out-of-sample distribution of daily return for SH000001, which is fitted by the 30min-based NIG-DCS model. It can be seen from the figure that the distribution has obvious sharp peaks and heavy tails. The out-of-sample daily VaR forecasts in our study are essentially the quantiles on the right tail of the fitted distribution.

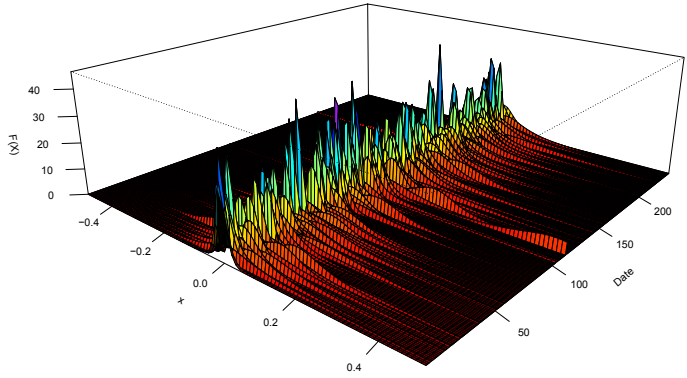

Figure 4.The fitted out-of-sample distribution of daily return,SH00001.

Hansen et al. (2012) proved that GARCH model integrated with realized measures of volatility can improve the empirical fitting efficiency compared to standard GARCH models based on daily returns only. Such models can be called RGARCH models and can extend to RGARCH-SSTD (skew-student distribution), RGARCH-GED (generalized error distribution), RGARCH-RV (realized volatility) and RGARCH-RRV (realized range-based volatility) according to different settings of volatility. We consider the above four RGARCH framework as the benchmarks and we will compare the performance of out-of-sample daily VaR forecasts that produced by 30min-NIG-DCS model and benchmark models, for SH00001 and SZZ399001 respectively.

Similarly, we only list part of results in Table 5 and Table 6, and the complete content of backtesting and MCS are in Appendix A. Table 5 implies that, when $\alpha = 0.99$, the performance of NIG-DCS-VaR are slightly inferior to other RGARCH-VaR. But at other levels, the NIG-DCS model based on 30min-HF data is better than other RGARCH models in terms of VaR coverage and independence，and it always ranks among the top three in the MCS test. Especially when $\alpha = \{0.93,0.94,0.95,0.96\}$, our model beats other RGARCH models and stays in the first place. In addition, the results reveal that when the distribution is not specified as GED, the RGARCH models with RRV as the realized measure outdo those RV-based. In RGARCH models, RGARCH-SSTD-RRV is considered as the best benchmarks and it outperforms our model when the risk level is very high.

Table 5. Back testing results of out-of-sample daily VaR forecasts, SH000001.

| | Alpha | LRUC statistics | LRCC statistics | DQ statistics | MCS ($\alpha$ = 0.15) | Alpha | LRUC statistics | LRCC statistics | DQ statistics | MCS($\alpha$ = 0.15) |
|---|---|---|---|---|---|---|---|---|---|---|
| NIG-DCS | | 0.28 (0.59) | 3.85 (0.15) | 13.88 (0.05) | 1 | | 0.37 (0.54) | 4.9 (0.09) | **14.97 (0.04*)** | 2 |
| RGARCH-SSTD-RV | | 1.41 (0.24) | 3.4 (0.18) | **17.95 (0.01*)** | 5 | | **4.99 (0.03*)** | 5.04 (0.08) | **22.86 (0**)** | 4 |
| RGARCH-GED-RV | | 1.41 (0.24) | 3.4 (0.18) | **17.99 (0.01*)** | 6 | | **4.99 (0.03*)** | 5.04 (0.08) | **22.89 (0**)** | 3 |
| RGARCH-NIG-RV | 0.93 | 1.41 (0.24) | 3.4 (0.18) | 17.94 (0.01*) | 4 | 0.97 | 3.71 (0.05) | 3.84 (0.15) | **22.1 (0**)** | 5 |
| RGARCH-SSTD-RRV | | 1.41 (0.24) | 3.4 (0.18) | 12.61 (0.08) | 2 | | 3.71 (0.05) | 3.84 (0.15) | 12.97 (0.07) | 1 |
| RGARCH-GED-RRV | | 1.41 (0.24) | 3.4 (0.18) | 12.21 (0.09) | 7 | | **4.99 (0.03*)** | **6.46 (0.04*)** | **18.35 (0.01*)** | 7 |
| RGARCH-NIG-RRV | | 0.91 (0.34) | 3.4 (0.18) | 14.49 (0.04) | 3 | | 3.71 (0.05) | 3.84 (0.15) | 12.97 (0.07) | 6 |
| | | | | | 0.34 | | | | | 0.36 |
| NIG-DCS | | 0.13 (0.72) | 3.15 (0.21) | **16.19 (0.02*)** | 1 | | 3.73 (0.05) | 6.11 (0.05) | **25.91 (0**)** | 5 |
| RGARCH-SSTD-RV | | 3.44 (0.06) | 4.9 (0.09) | **21.07 (0**)** | 3 | | 2.08 (0.15) | 2.29 (0.32) | **20.26 (0.01*)** | 4 |
| RGARCH-GED-RV | | 3.44 (0.06) | 4.9 (0.09) | **21.12 (0**)** | 5 | | 3.73 (0.05) | 6.11 (0.05) | **41.11 (0***)** | 6 |
| RGARCH-NIG-RV | 0.95 | 3.44 (0.06) | 4.9 (0.09) | **21.07 (0**)** | 4 | 0.99 | 2.08 (0.15) | 2.29 (0.32) | **20.26 (0.01*)** | 2 |
| RGARCH-SSTD-RRV | | **5.55 (0.02*)** | **8.04 (0.02*)** | **25.41 (0**)** | 2 | | 0.84 (0.36) | 0.98 (0.61) | 1.73 (0.97) | 1 |
| RGARCH-GED-RRV | | **5.55 (0.02*)** | **8.04 (0.02*)** | **24.27 (0**)** | 7 | | **8.01 (0**)** | **9.35 (0.01*)** | **40.83 (0***)** | 7 |
| RGARCH-NIG-RRV | | **5.55 (0.02*)** | **8.04 (0.02*)** | **25.43 (0**)** | 6 | | 2.08 (0.15) | 2.29 (0.32) | **28.49 (0**)** | 3 |
| | | | | | 0.38 | | | | | 0.35 |

Note: ∗, ∗∗, and ∗∗∗ represent statistical significance levels of 5%, 1%, and .1%, respectively. The rank tells the superiority of these four models under a default level of $\alpha$ and p-value helps to prove the non-rejection of this superiority. The p-value of the relevant test is indicated by the value in parentheses. Bold text indicates rejections at the * probability level.

Table 6 lists the test results of SZ399001. Compared with RGARCH, the NIG-DCS-VaR based on the empirical data of SZ3990001 show a more obvious advantage. Especially when the risk level increases, such as α>0.93, 30min-NIG-DCS model always ranks first in MCS test, and it can pass the LRUC, LRCC, and DQ tests with absolute advantage. Different from SH000001, RGARCH-GED-RV has the best performance among RGARCH models, followed by RGARCH-NIG-RV. And the results implies that the models using RV as the realized volatility is better than those RRV-based.

Table 6. Back testing results of out-of-sample daily VaR forecasts, SZ399001.

| | Alpha | LRuc statistics | LRcc statistics | DQ statistics | MCS ($\alpha$ = 0.15) | Alpha | LRuc statistics | LRcc statistics | DQ statistics | MCS ($\alpha$ = 0.15) |
|---|---|---|---|---|---|---|---|---|---|---|

| Model | | | | | | | | | | |
|---|---|---|---|---|---|---|---|---|---|---|
| NIG-DCS | 0.93 | 0.22 (0.64) | 1.69 (0.43) | 12.54 (0.08) | 2 | 0.97 | 0.37 (0.54) | 1.34 (0.51) | 5.5 (0.6) | 1 |
| RGARCH-SSTD-RV | | 0.91 (0.34) | 1.71 (0.43) | 8.4 (0.3) | 4 | | 3.71 (0.05) | 3.84 (0.15) | **38.36 (0**)** | 4 |
| RGARCH-GED-RV | | 0.91 (0.34) | 1.71 (0.43) | 8.52 (0.29) | 1 | | 3.71 (0.05) | 3.84 (0.15) | **38.68 (0**)** | 2 |
| RGARCH-NIG-RV | | 0.91 (0.34) | 1.71 (0.43) | 8.43 (0.3) | 3 | | 3.71 (0.05) | 3.84 (0.15) | **38.41 (0**)** | 3 |
| RGARCH-SSTD-RRV | | 1.41 (0.24) | 3.4 (0.18) | 10.9 (0.14) | 5 | | **4.99 (0.03*)** | 6.46 (0.04) | **38.84 (0**)** | 5 |
| RGARCH-GED-RRV | | 2 (0.16) | 3.57 (0.17) | 11.41 (0.12) | 7 | | **4.99 (0.03*)** | 6.46 (0.04) | **38.89 (0**)** | 7 |
| RGARCH-NIG-RRV | | 1.41 (0.24) | 3.4 (0.18) | 10.9 (0.14) | 6 | | **4.99 (0.03*)** | 6.46 (0.04) | **38.82 (0**)** | 6 |
| | | | | | 0.37 | | | | | 0.37 |
| NIG-DCS | 0.95 | 0.63 (0.43) | 0.64 (0.73) | 6.39 (0.5) | 1 | 0.99 | 2.08 (0.15) | 2.29 (0.32) | 12.53 (0.08) | 1 |
| RGARCH-SSTD-RV | | 2.55 (0.11) | 2.89 (0.24) | **14.87 (0.04*)** | 4 | | 3.73 (0.05) | 4.03 (0.13) | **25.71 (0**)** | 4 |
| RGARCH-GED-RV | | 2.55 (0.11) | 2.89 (0.24) | **14.99 (0.04*)** | 2 | | **5.72 (0.02*)** | 6.14 (0.05) | **35.24 (0**)** | 2 |
| RGARCH-NIG-RV | | 2.55 (0.11) | 2.89 (0.24) | **14.89 (0.04*)** | 3 | | 3.73 (0.05) | 4.03 (0.13) | **25.87 (0**)** | 3 |
| RGARCH-SSTD-RRV | | 2.55 (0.11) | 2.89 (0.24) | 13.44 (0.06) | 5 | | **13.33 (0**)** | **14 (0**)** | **65.95 (0***)** | 5 |
| RGARCH-GED-RRV | | 2.55 (0.11) | 2.89 (0.24) | 13.48 (0.06) | 7 | | **16.32 (0**)** | **16.76 (0**)** | **81.23 (0***)** | 7 |
| RGARCH-NIG-RRV | | 2.55 (0.11) | 2.89 (0.24) | 13.44 (0.06) | 6 | | **13.33 (0**)** | **14 (0**)** | **65.99 (0***)** | 6 |
| | | | | | 0.38 | | | | | 0.38 |

Note: ∗, ∗∗, and ∗∗∗ represent statistical significance levels of 5%, 1%, and .1%, respectively. The rank tells the superiority of these four models under a default level of $\alpha$ and p-value helps to prove the non-rejection of this superiority. The p-value of the relevant test is indicated by the value in parentheses. Bold text indicates rejections at the * probability level.

When $\alpha = \{0.94, 0.95, 0.96\}$, the out-of-sample NIG-DCS-VaR forecasts of SH000001 and SZ399001 have outstanding performance in contrast with RGARCH-VaR forecasts. Therefore, under these three risk levels, the VaR sequences generated by our model and benchmark models are visually compared in Figure 5. As is shown, when there is a cluster of extreme returns, for example, in mid-June and late August, NIG-DCS-VaR can sensitively capture the fluctuation of returns and cover the real return more accurately due to its unique DCS mechanism. The scores of the out-of-sample DCS models also passed the LM test. It is believed that there is no significant serial autocorrelation in the scores generated by NIG-DCS models based on the 30min returns. See for more details of the diagnosis from Appendix A.

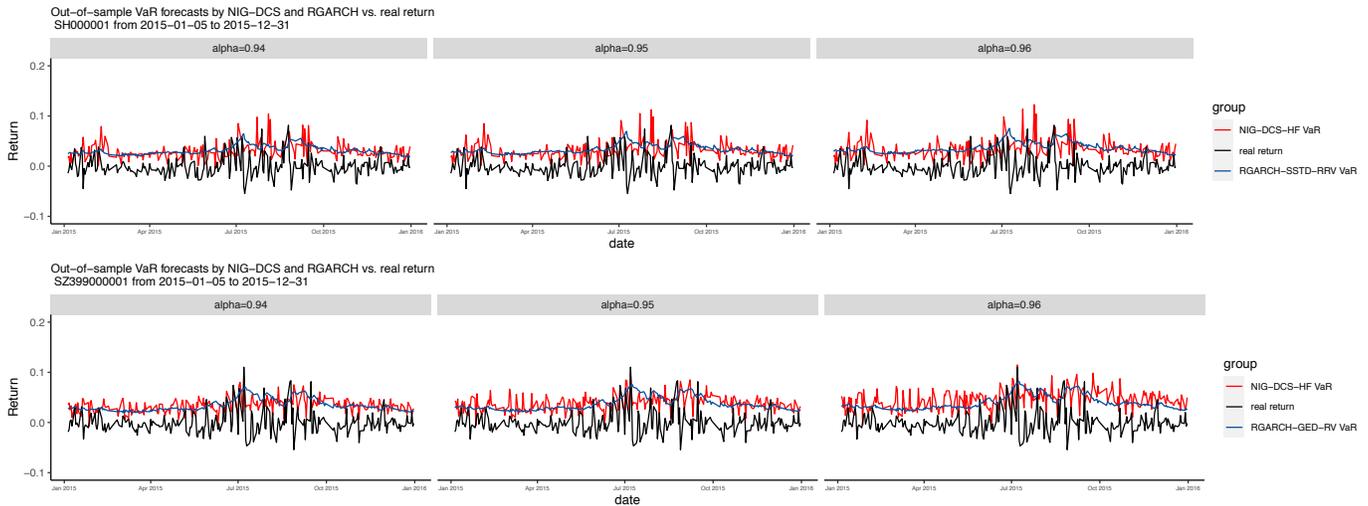

Figure 5. Comparison of out-of-sample VaR forecasts between 30min-NIG-DCS based and RGARCH based. The data is from January 5 to December 31 in 2015. The black line represents the real return, and blue line represents the out-of-sample VaR forecasts produced by RGARCH models.

It is revealed from out-of-sample results that the novel model proposed in this study could indeed improve the VaR measurement to a certain extent. The evidence to support it is that when $\alpha$ is in the range of 0.9-0.99, the NIG-DCS model based on HF data can provide more accurate VaR forecasts at most risk levels in contrast with the realized GARCH models. Hence, the NIG-DCS-VaR model built on a certain frequency of intraday returns makes a reliable contribution to financial risk management. And it may reduce the possibility of underestimating or overestimating the risk in financial supervision.

## 5. Conclusion

Encouraged by the literary fact that high-frequency data such as intraday returns contribute to financial risk management, we propose the NIG-DCS-VaR model in order to improve daily VaR evaluation by explicitly incorporating intraday returns. In relative research field, in order to make use of high-frequency data, realized volatility is commonly constructed based on intraday returns to well capture the noise of the market. Nevertheless, the realized-volatility-based approach often fails to fit the distribution of daily returns appropriately because the assumptions about the dynamic realized volatility and the form of distribution may not be consistent with the reality. Motivated by this situation, the NIG-DCS-VaR is developed to generate a forecast of daily return directly using intraday return. We introduce a parametric method and a non-parametric bootstrap method for combining the information that intraday returns offer. The parametric method takes advantage of the semi-additivity of the NIG distribution, and can be extended to various assets due to its strong tractability. We conducted an empirical analysis using two main indexes of the Chinese stock market, and a variety of backtesting approaches as well as the model confidence set approach prove that VaR forecasting performance of NIG-DCS is generally better than that of realized GARCH models. Especially when the risk level is relatively high, NIG-DCS-VaR beats RGARCH-VaR in terms of coverage ability and independence.

To sum up, our study contributes to enriching the theoretical understanding of VaR in several ways. First, it further emphasizes the effectiveness of intraday information in predicting daily VaR. Accurately estimating the distribution of intraday return is likely to be another key point for improving risk management, besides capturing intraday volatility. Second, NIG-DCS-VaR provides a basic dynamic frame that can be extended to other appropriate distributions specified for return. Third, the backtesting and MCS results demonstrate that NIG-DCS-VaR model can improve coverage ability, reduce the forecast error, and enhance statistical reliability in market risk management. In addition, we also found that although NIG distribution provides a convenient way of estimating daily return incorporating intraday returns due to its special property, it is not as effective as some special distributions, such as the Pareto distribution, in fitting returns with an extremely heavy tail. We will explore more possible optimization models under this framework, e.g., considering the combination of Weibull-Pareto distribution with DCS mechanism in our future work, to further promote the risk management capability of the VaR model.

**Declaration of interests**

The authors report no conflicts of interest. The authors alone are responsible for the content and writing of the paper.

Appendix A

Table A1. Back testing results of in-sample daily VaR estimates, SH000001.

| Model | Alpha | LRuc statistics | LRcc statistics | DQ statistics | MCS rank ($\alpha$ = 0.15) | Alpha | LRuc statistics | LRcc statistics | DQ statistics | MCS rank ($\alpha$ = 0.15) |
|---|---|---|---|---|---|---|---|---|---|---|
| VaR-day | 0.9 | **6.14 (0.01*)** | **8.44 (0.01*)** | 8.38 (0.3) | 5 | 0.95 | 1.23 (0.27) | 2.07 (0.35) | 4.14 (0.76) | 4 |
| VaR-40minhq | | 3.32 (0.07) | 6.49 (0.04) | 8.52 (0.29) | 2 | | 0.29 (0.59) | 1.48 (0.48) | 3.3 (0.86) | 1 |
| VaR-30minhq | | **5.09 (0.02*)** | 7.67 (0.02) | 8.41 (0.3) | 3 | | 0.68 (0.41) | 1.68 (0.43) | 3.21 (0.87) | 2 |
| VaR-20minhq | | 3.32 (0.07) | 3.51 (0.17) | 7.17 (0.41) | 1 | | 1.23 (0.27) | 2.07 (0.35) | 3.72 (0.81) | 3 |
| VaR-10minhq | | 1.42 (0.23) | 3.54 (0.17) | **16.58 (0.02*)** | 4 | | 0.07 (0.79) | 2.13 (0.34) | **15.86 (0.03*)** | 5 |
| | | | | | (0.54) | | | | | (0.33) |
| VaR-day | 0.91 | **5.69 (0.02*)** | **7.5 (0.02*)** | 8.31 (0.31) | 5 | 0.96 | 0.85 (0.36) | 1.41 (0.49) | 4.33 (0.74) | 4 |
| VaR-40minhq | | 2.17 (0.14) | 5.04 (0.08) | 6.16 (0.52) | 1 | | 0.09 (0.77) | 0.93 (0.63) | 3.15 (0.87) | 2 |
| VaR-30minhq | | **4.63 (0.03*)** | 6.68 (0.04) | 7.51 (0.38) | 3 | | 0.09 (0.77) | 0.93 (0.63) | 2.87 (0.9) | 1 |
| VaR-20minhq | | **4.63 (0.03*)** | 6.68 (0.04) | 6.44 (0.49) | 2 | | 0.85 (0.36) | 1.41 (0.49) | 3.91 (0.79) | 3 |
| VaR-10minhq | | 0.68 (0.41) | 1.54 (0.46) | 12.94 (0.07) | 4 | | 0.08 (0.77) | 2.62 (0.27) | 11.87 (0.1) | 5 |
| | | | | | (0.28) | | | | | (0.45) |
| VaR-day | 0.92 | 3.25 (0.07) | 5.06 (0.08) | 6.46 (0.49) | 5 | 0.97 | 1.17 (0.28) | 1.41 (0.49) | 2.44 (0.93) | 5 |
| VaR-40minhq | | 1.77 (0.18) | 4.08 (0.13) | 6.63 (0.47) | 1 | | 0.5 (0.48) | 0.83 (0.66) | 1.88 (0.97) | 2 |
| VaR-30minhq | | 3.25 (0.07) | 5.06 (0.08) | 6.11 (0.53) | 2 | | 0.5 (0.48) | 0.83 (0.66) | 1.88 (0.97) | 1 |
| VaR-20minhq | | 4.19 (0.04) | 5.78 (0.06) | 6.27 (0.51) | 3 | | 0.12 (0.73) | 0.56 (0.76) | 4.06 (0.77) | 3 |
| VaR-10minhq | | 0.19 (0.67) | 1.32 (0.52) | 12.99 (0.07) | 4 | | 0.43 (0.51) | 1.12 (0.57) | 5.73 (0.57) | 4 |
| | | | | | (0.42) | | | | | (0.56) |
| VaR-day | 0.93 | 2.82 (0.09) | 4.2 (0.12) | 5.55 (0.59) | 4 | 0.98 | 0.77 (0.38) | 0.88 (0.64) | 1.83 (0.97) | 5 |
| VaR-40minhq | | 0.48 (0.49) | 2.79 (0.25) | 5.9 (0.55) | 1 | | 0.18 (0.67) | 0.35 (0.84) | 1.32 (0.99) | 2 |
| VaR-30minhq | | 2.03 (0.15) | 3.62 (0.16) | 4.8 (0.68) | 2 | | 0.18 (0.67) | 0.35 (0.84) | 1.32 (0.99) | 1 |
| VaR-20minhq | | 3.76 (0.05) | 4.94 (0.08) | 5.6 (0.59) | 3 | | 0 (1) | 0.25 (0.88) | 7.69 (0.36) | 3 |
| VaR-10minhq | | 0.48 (0.49) | 1.21 (0.55) | 11.71 (0.11) | 5 | | 0 (1) | 0.25 (0.88) | 1.71 (0.97) | 4 |
| | | | | | (0.66) | | | | | (0.42) |
| VaR-day | 0.94 | 2.4 (0.12) | 3.4 (0.18) | 5 (0.66) | 4 | 0.99 | 0.3 (0.58) | 0.41 (0.81) | 0.69 (1) | 4 |
| VaR-40minhq | | 0.25 (0.62) | 2.06 (0.36) | 4.33 (0.74) | 1 | | 0.3 (0.58) | 0.41 (0.81) | 0.64 (1) | 2 |
| VaR-30minhq | | 1.63 (0.2) | 2.81 (0.25) | 4.04 (0.78) | 2 | | 0.3 (0.58) | 0.41 (0.81) | 0.64 (1) | 1 |
| VaR-20minhq | | 2.4 (0.12) | 3.4 (0.18) | 4.49 (0.72) | 3 | | 0.32 (0.58) | 0.41 (0.81) | 0.64 (1) | 3 |
| VaR-10minhq | | 0.25 (0.62) | 1.53 (0.46) | **14.73 (0.04*)** | 5 | | 1.12 (0.29) | 1.29 (0.52) | 3.7 (0.81) | 5 |
| | | | | | (0.72) | | | | | (0.35) |

Note: ∗, ∗∗, and ∗∗∗ represent statistical significance levels of 5%, 1%, and .1%, respectively. The rank tells the superiority of these four models under a default level of $\alpha$ and p-value helps to prove the non-rejection of this

superiority. The p-value of the relevant test is indicated by the value in parentheses. Bold text indicates rejections at the * probability level.

Table A2. Back testing results of in-sample daily VaR estimates, SZ399001.

| Model | Alpha | LRuc statistics | LRcc statistics | DQ statistics | MCS rank ($\alpha = 0.15$) | Alpha | LRuc statistics | LRcc statistics | DQ statistics | MCS rank ($\alpha = 0.15$) |
|---|---|---|---|---|---|---|---|---|---|---|
| VaR-day | 0.9 | 3.34 (0.07) | 3.87 (0.14) | 8.71 (0.27) | 2 | 0.95 | 10.26 (0**) | 10.26 (0.01*) | 5.05 (0.65) | 4 |
| VaR-40minhq | | 10.25 (0**) | 10.34 (0.01*) | 7.09 (0.42) | 4 | | 4.95 (0.03*) | 4.97 (0.08) | 4.33 (0.74) | 3 |
| VaR-30minhq | | 0.11 (0.74) | 0.66 (0.72) | 20.33 (0**) | 1 | | 0 (1) | 1.51 (0.47) | 20.37 (0**) | 1 |
| VaR-20minhq | | 14.27 (0**) | 14.29 (0**) | 8.76 (0.27) | 5 | | 10.26 (0**) | 10.26 (0.01*) | 5.05 (0.65) | 5 |
| VaR-10minhq | | 7.31 (0.01*) | 7.49 (0.02*) | 14.92 (0.04*) | 3 | | 0.98 (0.32) | 1.16 (0.56) | 17.26 (0.02*) | 2 |
| | | | | | (0.29) | | | | | (0.44) |
| VaR-day | 0.91 | 3.78 (0.05) | 4.12 (0.13) | 8.49 (0.29) | 2 | 0.96 | 8.16 (0**) | 8.16 (0.02*) | 4 (0.78) | 4 |
| VaR-40minhq | | 12.29 (0**) | 12.31 (0**) | 7.94 (0.34) | 4 | | 3.32 (0.07) | 3.34 (0.19) | 3.77 (0.81) | 3 |
| VaR-30minhq | | 0.13 (0.72) | 0.33 (0.85) | 22.11 (0**) | 1 | | 0.24 (0.62) | 1.75 (0.42) | 14.47 (0.05) | 1 |
| VaR-20minhq | | 18.86 (0**) | 18.86 (0**) | 9.49 (0.22) | 5 | | 8.16 (0**) | 8.16 (0.02*) | 4 (0.78) | 5 |
| VaR-10minhq | | 5.8 (0.02*) | 5.98 (0.05) | 14.65 (0.04*) | 3 | | 0.28 (0.59) | 0.47 (0.79) | 20.12 (0.01*) | 2 |
| | | | | | (0.34) | | | | | (0.49) |
| VaR-day | 0.92 | 4.38 (0.04*) | 4.57 (0.1) | 7.29 (0.4) | 2 | 0.97 | 6.09 (0.01*) | 6.09 (0.05) | 2.97 (0.89) | 4 |
| VaR-40minhq | | 10.36 (0**) | 10.38 (0.01*) | 6.93 (0.44) | 4 | | 1.84 (0.17) | 1.86 (0.39) | 3.6 (0.82) | 3 |
| VaR-30minhq | | 0 (1) | 0.2 (0.9) | 24.1 (0**) | 1 | | 1.15 (0.28) | 2.66 (0.26) | 11.85 (0.1) | 1 |
| VaR-20minhq | | 16.68 (0**) | 16.68 (0**) | 8.35 (0.3) | 5 | | 6.09 (0.01*) | 6.09 (0.05) | 2.97 (0.89) | 5 |
| VaR-10minhq | | 4.38 (0.04*) | 4.57 (0.1) | 14.63 (0.04*) | 3 | | 0 (1) | 0.19 (0.91) | 25.58 (0**) | 2 |
| | | | | | (0.35) | | | | | (0.38) |
| VaR-day | 0.93 | 5.25 (0.02*) | 5.34 (0.07) | 5.98 (0.54) | 3 | 0.98 | 4.04 (0.04*) | 4.04 (0.13) | 1.96 (0.96) | 4 |
| VaR-40minhq | | 8.49 (0**) | 8.51 (0.01*) | 5.97 (0.54) | 4 | | 4.04 (0.04*) | 4.04 (0.13) | 1.96 (0.96) | 3 |
| VaR-30minhq | | 0 (1) | 0.48 (0.79) | 15.18 (0.03*) | 1 | | 0.44 (0.51) | 4.07 (0.13) | 10.74 (0.14) | 1 |
| VaR-20minhq | | 14.51 (0**) | 14.51 (0**) | 7.23 (0.41) | 5 | | 4.04 (0.04*) | 4.04 (0.13) | 1.96 (0.96) | 5 |
| VaR-10minhq | | 3.09 (0.08) | 3.27 (0.19) | 14.93 (0.04*) | 2 | | 0 (1) | 0.08 (0.96) | 25.53 (0**) | 2 |
| | | | | | (0.55) | | | | | (0.42) |
| VaR-day | 0.94 | 3.77 (0.05) | 3.86 (0.15) | 5.36 (0.62) | 3 | 0.99 | 2.01 (0.16) | 2.01 (0.37) | 0.97 (1) | 4 |
| VaR-40minhq | | 6.68 (0.01) | 6.7 (0.04*) | 5.09 (0.65) | 4 | | 2.01 (0.16) | 2.01 (0.37) | 0.97 (1) | 3 |

| | | | | | MCS | | | | | MCS |
|---|---|---|---|---|---|---|---|---|---|---|---|
| | Alpha | LRuc statistics | LRcc statistics | DQ statistics | ($\alpha$ = 0.15) | Alpha | LRuc statistics | LRcc statistics | DQ statistics | ($\alpha$ = 0.15) |
| VaR-30minhq | | 0 (1) | 0.9 (0.64) | 16.53 (0.02*) | 1 | | 0 (1) | 0.02 (0.99) | 12.81 (0.06) | 1 |
| VaR-20minhq | | 12.38 (0**) | 12.38 (0**) | 6.13 (0.52) | 5 | | 2.01 (0.16) | 2.01 (0.37) | 0.97 (1) | 5 |
| VaR-10minhq | | 1.94 (0.16) | 2.12 (0.35) | 15.72 (0.03*) | 2 | | 0.78 (0.38) | 0.87 (0.65) | 49.64 (0**) | 2 |
| | | | | | (0.52) | | | | | (0.45) |

Note: *, **, and *** represent statistical significance levels of 5%, 1%, and .1%, respectively. The rank tells the superiority of these four models under a default level of $\alpha$ and p-value helps to prove the non-rejection of this superiority. The p-value of the relevant test is indicated by the value in parentheses. Bold text indicates rejections at the * probability level.

Table A3. Back testing results of out-of-sample daily VaR forecasts, SH000001.

| | Alpha | LRuc statistics | LRcc statistics | DQ statistics | MCS ($\alpha$ = 0.15) | Alpha | LRuc statistics | LRcc statistics | DQ statistics | MCS ($\alpha$ = 0.15) |
|---|---|---|---|---|---|---|---|---|---|---|
| NIG-DCS | | 0.55 (0.46) | 3.04 (0.22) | 12.81 (0.08) | 3 | | 0.13 (0.72) | 3.15 (0.21) | **16.19 (0.02*)** | 1 |
| RGARCH-SSTD-RV | | 0.3 (0.58) | 1.77 (0.41) | 8.15 (0.32) | 5 | | 3.44 (0.06) | 4.9 (0.09) | **21.07 (0**)** | 3 |
| RGARCH-GED-RV | | 0.3 (0.58) | 1.77 (0.41) | 8.16 (0.32) | 6 | | 3.44 (0.06) | 4.9 (0.09) | **21.12 (0**)** | 5 |
| RGARCH-NIG-RV | 0.9 | 0.3 (0.58) | 1.77 (0.41) | 8.15 (0.32) | 4 | 0.95 | 3.44 (0.06) | 4.9 (0.09) | **21.07 (0**)** | 4 |
| RGARCH-SSTD-RRV | | 0.57 (0.45) | 1.68 (0.43) | 12.62 (0.08) | 2 | | **5.55 (0.02*)** | **8.04 (0.02*)** | **25.41 (0**)** | 2 |
| RGARCH-GED-RRV | | 0.57 (0.45) | 1.68 (0.43) | 11.74 (0.11) | 7 | | **5.55 (0.02*)** | **8.04 (0.02*)** | **24.27 (0**)** | 7 |
| RGARCH-NIG-RRV | | 0.57 (0.45) | 1.68 (0.43) | 12.64 (0.08) | 1 | | **5.55 (0.02*)** | **8.04 (0.02*)** | **25.43 (0**)** | 6 |
| | | | | | 0.39 | | | | | 0.38 |
| NIG-DCS | | 0.83 (0.36) | 2.72 (0.26) | 13.33 (0.06) | 2 | | 0.01 (0.94) | 3.72 (0.16) | 12.58 (0.06) | 1 |
| RGARCH-SSTD-RV | | 0.44 (0.5) | 2.81 (0.25) | 12.52 (0.08) | 5 | | **4.61 (0.03*)** | **4.65 (0.1)** | **24.81 (0**)** | 4 |
| RGARCH-GED-RV | | 0.44 (0.5) | 2.81 (0.25) | 12.52 (0.08) | 6 | | **4.61 (0.03*)** | **4.65 (0.1)** | **24.89 (0**)** | 3 |
| RGARCH-NIG-RV | 0.91 | 0.44 (0.5) | 2.81 (0.25) | 12.51 (0.08) | 4 | 0.96 | **4.61 (0.03*)** | **4.65 (0.1)** | **24.81 (0**)** | 5 |
| RGARCH-SSTD-RRV | | 1.19 (0.27) | 2.66 (0.26) | 11.9 (0.1) | 1 | | 3.51 (0.06) | 4.3 (0.12) | **15.48 (0.03*)** | 2 |
| RGARCH-GED-RRV | | 0.78 (0.38) | 2.67 (0.26) | 9.77 (0.2) | 7 | | 2.53 (0.11) | 3.64 (0.16) | 10.31 (0.17) | 7 |
| RGARCH-NIG-RRV | | 1.19 (0.27) | 2.66 (0.26) | 11.92 (0.1) | 3 | | 2.53 (0.11) | 3.64 (0.16) | 9.74 (0.2) | 6 |
| | | | | | 0.42 | | | | | 0.36 |
| NIG-DCS | | 0.37 (0.54) | 2.75 (0.25) | 11.19 (0.13) | 1 | | 0.37 (0.54) | 4.9 (0.09) | **14.97 (0.04*)** | 2 |
| RGARCH-SSTD-RV | | 1.05 (0.31) | 3.95 (0.14) | **16.35 (0.02*)** | 5 | | **4.99 (0.03*)** | 5.04 (0.08) | **22.86 (0**)** | 4 |
| RGARCH-GED-RV | | 1.05 (0.31) | 3.95 (0.14) | **16.34 (0.02*)** | 7 | | **4.99 (0.03*)** | 5.04 (0.08) | **22.89 (0**)** | 3 |
| RGARCH-NIG-RV | 0.92 | 1.05 (0.31) | 3.95 (0.14) | **16.33 (0.02*)** | 3 | 0.97 | 3.71 (0.05) | 3.84 (0.15) | **22.1 (0**)** | 5 |
| RGARCH-SSTD-RRV | | 0.64 (0.42) | 4.15 (0.13) | 14.13 (0.05) | 2 | | 3.71 (0.05) | 3.84 (0.15) | 12.97 (0.07) | 1 |
| RGARCH-GED-RRV | | 0.64 (0.42) | 4.15 (0.13) | 13.77 (0.06) | 4 | | **4.99 (0.03*)** | **6.46 (0.04*)** | **18.35 (0.01*)** | 7 |
| RGARCH-NIG-RRV | | 0.64 (0.42) | 4.15 (0.13) | 14.14 (0.05) | 6 | | 3.71 (0.05) | 3.84 (0.15) | 12.97 (0.07) | 6 |
| | | | | | 0.37 | | | | | 0.36 |
| NIG-DCS | 0.93 | 0.28 (0.59) | 3.85 (0.15) | 13.88 (0.05) | 1 | 0.98 | 0.24 (0.62) | 2.62 (0.27) | 10.93 (0.14) | 1 |
| RGARCH-SSTD-RV | | 1.41 (0.24) | 3.4 (0.18) | **17.95 (0.01*)** | 5 | | **4.22 (0.04*)** | 4.89 (0.09) | **23.53 (0**)** | 3 |
| RGARCH-GED-RV | | 1.41 (0.24) | 3.4 (0.18) | **17.99 (0.01*)** | 6 | | **4.22 (0.04*)** | 4.89 (0.09) | **23.57 (0**)** | 2 |
| RGARCH-NIG-RV | | 1.41 (0.24) | 3.4 (0.18) | 17.94 (0.01*) | 4 | | **4.22 (0.04*)** | 4.89 (0.09) | **23.54 (0**)** | 5 |
| RGARCH-SSTD-RRV | | 1.41 (0.24) | 3.4 (0.18) | 12.61 (0.08) | 2 | | **5.8 (0.02*)** | 6.24 (0.04*) | **26.39 (0**)** | 4 |
| RGARCH-GED-RRV | | 1.41 (0.24) | 3.4 (0.18) | 12.21 (0.09) | 7 | | **5.8 (0.02*)** | 6.24 (0.04*) | **29.55 (0**)** | 7 |
| RGARCH-NIG-RRV | | 0.91 (0.34) | 3.4 (0.18) | 14.49 (0.04) | 3 | | **5.8 (0.02*)** | 6.24 (0.04*) | **26.39 (0**)** | 6 |
| | | | | | 0.34 | | | | | 0.36 |
| NIG-DCS | | 0.01 (0.92) | 3.58 (0.17) | 15 (0.06) | 1 | | 3.73 (0.05) | 6.11 (0.05) | **25.91 (0**)** | 5 |
| RGARCH-SSTD-RV | | 3.44 (0.06) | 5.44 (0.07) | 23 (0) | 4 | | 2.08 (0.15) | 2.29 (0.32) | **20.26 (0.01*)** | 4 |
| RGARCH-GED-RV | | 3.44 (0.06) | 5.44 (0.07) | 23.04 (0) | 6 | | 3.73 (0.05) | 6.11 (0.05) | **41.11 (0***)** | 6 |
| RGARCH-NIG-RV | 0.94 | 3.44 (0.06) | 5.44 (0.07) | 22.99 (0) | 5 | 0.99 | 2.08 (0.15) | 2.29 (0.32) | **20.26 (0.01*)** | 2 |
| RGARCH-SSTD-RRV | | 2.61 (0.11) | 5.1 (0.08) | 18.63 (0.01) | 2 | | 0.84 (0.36) | 0.98 (0.61) | 1.73 (0.97) | 1 |
| RGARCH-GED-RRV | | 2.61 (0.11) | 5.1 (0.08) | 17.66 (0.01) | 7 | | **8.01 (0**)** | **9.35 (0.01*)** | **40.83 (0***)** | 7 |
| RGARCH-NIG-RRV | | 2.61 (0.11) | 5.1 (0.08) | 18.65 (0.01) | 3 | | 2.08 (0.15) | 2.29 (0.32) | **28.49 (0**)** | 3 |
| | | | | | 0.36 | | | | | 0.35 |

Note: ∗, ∗∗, and ∗∗∗ represent statistical significance levels of 5%, 1%, and .1%, respectively. The rank tells the superiority of these four models under a default level of $\alpha$ and p-value helps to prove the non-rejection of this superiority. The p-value of the relevant test is indicated by the value in parentheses. Bold text indicates rejections at the * probability level.

Table A4. Back testing results of out-of-sample daily VaR forecasts, SZ399001.

| | Alpha | LRuc statistics | LRcc statistics | DQ statistics | MCS ($\alpha = 0.15$) | Alpha | LRuc statistics | LRcc statistics | DQ statistics | MCS ($\alpha = 0.15$) |
|---|---|---|---|---|---|---|---|---|---|---|
| NIG-DCS | 0.9 | 0.02 (0.9) | 0.1 (0.95) | 8.16 (0.32) | 2 | 0.95 | 0.63 (0.43) | 0.64 (0.73) | 6.39 (0.5) | 1 |
| RGARCH-SSTD-RV | | 0.02 (0.9) | 0.89 (0.64) | 4.67 (0.7) | 4 | | 2.55 (0.11) | 2.89 (0.24) | **14.87 (0.04*)** | 4 |
| RGARCH-GED-RV | | 0.02 (0.9) | 0.89 (0.64) | 4.73 (0.69) | 1 | | 2.55 (0.11) | 2.89 (0.24) | **14.99 (0.04*)** | 2 |
| RGARCH-NIG-RV | | 0.02 (0.9) | 0.89 (0.64) | 4.68 (0.7) | 3 | | 2.55 (0.11) | 2.89 (0.24) | **14.89 (0.04*)** | 3 |
| RGARCH-SSTD-RRV | | 0.11 (0.74) | 2 (0.37) | 9.7 (0.21) | 5 | | 2.55 (0.11) | 2.89 (0.24) | 13.44 (0.06) | 5 |
| RGARCH-GED-RRV | | 0.11 (0.74) | 2 (0.37) | 9.71 (0.21) | 7 | | 2.55 (0.11) | 2.89 (0.24) | 13.48 (0.06) | 7 |
| RGARCH-NIG-RRV | | 0.11 (0.74) | 2 (0.37) | 9.7 (0.21) | 6 | | 2.55 (0.11) | 2.89 (0.24) | 13.44 (0.06) | 6 |
| | | | | | 0.39 | | | | | 0.38 |
| NIG-DCS | 0.91 | 0.44 (0.5) | 0.53 (0.77) | 9.15 (0.24) | 2 | 0.96 | 0.5 (0.48) | 0.76 (0.68) | 8.24 (0.31) | 1 |
| RGARCH-SSTD-RV | | 0.05 (0.82) | 0.4 (0.82) | 3.41 (0.85) | 4 | | 3.51 (0.06) | 4.3 (0.12) | **23.9 (0**)** | 4 |
| RGARCH-GED-RV | | 0.2 (0.65) | 0.4 (0.82) | 3.96 (0.78) | 1 | | 3.51 (0.06) | 4.3 (0.12) | **24.07 (0**)** | 2 |
| RGARCH-NIG-RV | | 0.05 (0.82) | 0.4 (0.82) | 3.42 (0.84) | 3 | | 3.51 (0.06) | 4.3 (0.12) | **23.93 (0**)** | 3 |
| RGARCH-SSTD-RRV | | 0.2 (0.65) | 1.39 (0.5) | 10.44 (0.17) | 5 | | **5.85 (0.02*)** | 6.19 (0.05) | **20.32 (0**)** | 5 |
| RGARCH-GED-RRV | | 0.2 (0.65) | 1.39 (0.5) | 10.43 (0.17) | 7 | | **5.85 (0.02*)** | 6.19 (0.05) | **20.36 (0**)** | 7 |
| RGARCH-NIG-RRV | | 0.2 (0.65) | 1.39 (0.5) | 10.44 (0.16) | 6 | | **5.85 (0.02*)** | 6.19 (0.05) | **20.31 (0**)** | 6 |
| | | | | | 0.38 | | | | | 0.38 |
| NIG-DCS | 0.92 | 0.33 (0.57) | 0.88 (0.64) | 10.49 (0.16) | 4 | 0.97 | 0.37 (0.54) | 1.34 (0.51) | 5.5 (0.6) | 1 |
| RGARCH-SSTD-RV | | 0.12 (0.73) | 0.92 (0.63) | 6.65 (0.47) | 3 | | 3.71 (0.05) | 3.84 (0.15) | **38.36 (0**)** | 4 |
| RGARCH-GED-RV | | 0.12 (0.73) | 0.92 (0.63) | 6.76 (0.45) | 1 | | 3.71 (0.05) | 3.84 (0.15) | **38.68 (0**)** | 2 |
| RGARCH-NIG-RV | | 0.12 (0.73) | 0.92 (0.63) | 6.68 (0.46) | 2 | | 3.71 (0.05) | 3.84 (0.15) | **38.41 (0**)** | 3 |
| RGARCH-SSTD-RRV | | 0.64 (0.42) | 2.21 (0.33) | 8.75 (0.27) | 5 | | **4.99 (0.03*)** | 6.46 (0.04) | **38.84 (0**)** | 5 |
| RGARCH-GED-RRV | | 0.64 (0.42) | 2.21 (0.33) | 8.75 (0.27) | 7 | | **4.99 (0.03*)** | 6.46 (0.04) | **38.89 (0**)** | 7 |
| RGARCH-NIG-RRV | | 0.64 (0.42) | 2.21 (0.33) | 8.75 (0.27) | 6 | | **4.99 (0.03*)** | 6.46 (0.04) | **38.82 (0**)** | 6 |
| | | | | | 0.32 | | | | | 0.37 |
| NIG-DCS | 0.93 | 0.22 (0.64) | 1.69 (0.43) | 12.54 (0.08) | 2 | 0.98 | 1.71 (0.19) | 3.05 (0.22) | 10.02 (0.19) | 1 |
| RGARCH-SSTD-RV | | 0.91 (0.34) | 1.71 (0.43) | 8.4 (0.3) | 4 | | **7.57 (0.01*)** | **7.83 (0.02*)** | **34.93 (0**)** | 4 |
| RGARCH-GED-RV | | 0.91 (0.34) | 1.71 (0.43) | 8.52 (0.29) | 1 | | **7.57 (0.01*)** | **7.83 (0.02*)** | **35.26 (0**)** | 2 |
| RGARCH-NIG-RV | | 0.91 (0.34) | 1.71 (0.43) | 8.43 (0.3) | 3 | | **7.57 (0.01*)** | **7.83 (0.02*)** | **34.97 (0**)** | 3 |
| RGARCH-SSTD-RRV | | 1.41 (0.24) | 3.4 (0.18) | 10.9 (0.14) | 5 | | **9.51 (0**)** | **9.65 (0.01*)** | **61.37 (0**)** | 5 |
| RGARCH-GED-RRV | | 2 (0.16) | 3.57 (0.17) | 11.41 (0.12) | 7 | | **9.51 (0**)** | **9.65 (0.01*)** | **61.72 (0**)** | 7 |
| RGARCH-NIG-RRV | | 1.41 (0.24) | 3.4 (0.18) | 10.9 (0.14) | 6 | | **7.57 (0.01*)** | **7.83 (0.02*)** | **64.9 (0**)** | 6 |
| | | | | | 0.37 | | | | | 0.37 |
| NIG-DCS | 0.94 | 0.77 (0.38) | 1.11 (0.57) | 5.15 (0.64) | 1 | 0.99 | 2.08 (0.15) | 2.29 (0.32) | 12.53 (0.08) | 1 |
| RGARCH-SSTD-RV | | 1.27 (0.26) | 1.46 (0.48) | 10.2 (0.18) | 4 | | 3.73 (0.05) | 4.03 (0.13) | **25.71 (0**)** | 4 |
| RGARCH-GED-RV | | 0.77 (0.38) | 1.11 (0.57) | 10.97 (0.14) | 2 | | **5.72 (0.02*)** | 6.14 (0.05) | **35.24 (0**)** | 2 |
| RGARCH-NIG-RV | | 1.27 (0.26) | 1.46 (0.48) | 10.23 (0.18) | 3 | | 3.73 (0.05) | 4.03 (0.13) | **25.87 (0**)** | 3 |
| RGARCH-SSTD-RRV | | 1.89 (0.17) | 2.99 (0.22) | 11.56 (0.12) | 5 | | **13.33 (0**)** | **14 (0**)** | **65.95 (0***)** | 5 |
| RGARCH-GED-RRV | | 1.89 (0.17) | 4.94 (0.08) | **17.04 (0.02*)** | 7 | | **16.32 (0**)** | **16.76 (0**)** | **81.23 (0***)** | 7 |
| RGARCH-NIG-RRV | | 1.89 (0.17) | 2.99 (0.22) | 11.55 (0.12) | 6 | | **13.33 (0**)** | **14 (0**)** | **65.99 (0***)** | 6 |
| | | | | | 0.36 | | | | | 0.38 |

Note: ∗, ∗∗, and ∗∗∗ represent statistical significance levels of 5%, 1%, and .1%, respectively. The rank tells the superiority of these four models under a default level of $\alpha$ and p-value helps to prove the non-rejection of this superiority. The p-value of the relevant test is indicated by the value in parentheses. Bold text indicates rejections at the * probability level.

Table A5. P-values of LM test for in-sample parameter's scores of daily return.

| | mu | lamda | nu | theta |
|---|---|---|---|---|
| SH000001 | 0.13 | 0.38 | 0.17 | 0.1 |
| SZ399001 | 0.22 | 0.61 | 0.61 | 0.46 |

Note: a p-value greater than 0.05 indicates that the null hypothesis that the sequence is not self-correlated is not rejected.

Table A6. P-values of LM test for in-sample parameter's scores of 30min return.

| Period | SH000001-mu | SH000001-nv | SZ399001-mu | SZ399001-nv |
|---|---|---|---|---|
| 1 | 0.18 | 1 | 0.49 | 1 |
| 2 | 0.73 | 1 | 0.94 | 0.88 |
| 3 | 0.8 | 0.55 | 0.34 | 0.4 |
| 4 | 0.97 | 0.34 | 0.17 | 1 |
| 5 | 0.18 | 0.95 | 0.96 | 0.16 |
| 6 | 1 | 0.11 | 0.42 | 1 |
| 7 | 0.56 | 0.05 | 0.13 | 0.1 |
| 8 | 0.87 | 0.22 | 0.48 | 1 |

Note: a p-value greater than 0.05 indicates that the null hypothesis that the sequence is not self-correlated is not rejected.

Table A7. P-values of LM test for out-of-sample parameter's scores of daily return.

|  | mu | lamda | nu | theta |
|---|---|---|---|---|
| SH000001 | 1 | 1 | 0.28 | 1 |
| SZ399001 | 0.35 | 0.98 | 0.5 | 0.14 |

Note: a p-value greater than 0.05 indicates that the null hypothesis that the sequence is not self-correlated is not rejected.

Table A8. P-values of LM test for out-of-sample parameter's scores of 30min return.

|  | SH000001-mu | SH000001-nv | SZ399001-mu | SZ399001-nv |
|---|---|---|---|---|
| 1 | 0.88 | 0.3 | 0.32 | 0.23 |
| 2 | 0.28 | 0.94 | 0.25 | 0 |
| 3 | 0.77 | 0.1 | 1 | 0.21 |
| 4 | 0.37 | 0.01 | 1 | 0.12 |
| 5 | 1 | 0.96 | 0.95 | 0.55 |
| 6 | 0.81 | 0.34 | 0.89 | 1 |
| 7 | 0.41 | 0.39 | 0.97 | 0.57 |
| 8 | 0.09 | 0.93 | 0.82 | 0.84 |

Note: a p-value greater than 0.05 indicates that the null hypothesis that the sequence is not self-correlated is not rejected.